\documentclass[10pt,oneside,letterpaper]{article}
\usepackage{amssymb}
\usepackage{amsmath}
\usepackage[dvips]{graphicx}
\usepackage{setspace}

\begin{document}

\baselineskip=18pt
\numberwithin{equation}{section}
\allowdisplaybreaks  

\pagestyle{myheadings}

\thispagestyle{empty}

\vspace*{-2cm}
\begin{flushright}
{\tt hep-th/0705.0505}\\
\end{flushright}

\vspace*{2.5cm}
\begin{center}
 {\LARGE Thermal Evolution of the \\ Non Supersymmetric Metastable Vacua in \\ ${\cal N}=2$ $SU(2)$ SYM Softly Broken to ${\cal N}=1$\\}
 \vspace*{1.7cm}
 Eleni Katifori$^1$, Georgios Pastras$^1$\\
 \vspace*{1.0cm}
  $^1$ Jefferson Physical Laboratory, Harvard University, Cambridge, MA 02138,
USA\\
 \vspace*{0.8cm}
 {\tt pastras@fas.harvard.edu}
\end{center}
\vspace*{1.5cm}

\noindent

It has been shown that four dimensional ${\cal N}=2$ gauge theories, softly broken to ${\cal N}=1$ by a superpotential term, can accommodate metastable non-supersymmetric vacua in their moduli space. We study the $SU(2)$ theory at high temperatures in order to determine whether a cooling universe settles in the metastable vacuum at zero temperature. We show that the corrections to the free energy because of the BPS dyons are such that may destroy the existence of the metastable vacuum at high temperatures. Nevertheless we demonstrate the universe can settle in the metastable vacuum, provided that the following two conditions are hold: first the superpotential term is not arbitrarily small in comparison to the strong coupling scale of the gauge theory, and second the metastable vacuum lies in the strongly coupled region of the moduli space.

\newpage
\setcounter{page}{1}

\onehalfspacing

\section{Introduction}

Sypersymmetry breaking in a metastable vacuum is an appealing choice for constructing realistic models where sypersymmetry has to be broken. In a pioneering paper \cite{Intriligator:2006dd} it was shown that such vacua exist in the configuration space of simple ${\cal N}=1$ theories like SQCD with massive flavors. Since then these ideas have found fertile ground in field theory \cite{Kitano:2006wm, Banks:2006ma,
Schmaltz:2006qs, Dine:2006xt, Kitano:2006xg, Murayama:2006yf,
Csaki:2006wi, Intriligator:2007py} and string theory \cite{Franco:2006es, Ooguri:2006pj, Ooguri:2006bg, Franco:2006ht,
Bena:2006rg, Argurio:2006ny, Aganagic:2006ex, Giveon:2007fk,
Argurio:2007qk, Kawano:2007ru, Marsano:2007fe}, resulting in a lot of extensions and realizations.

An interesting question already raised in \cite{Intriligator:2006dd} is whether ${\cal N}=2$ theories softly broken to ${\cal N}=1$ can accommodate such vacua. ${\cal N}=2$ theories have moduli spaces, where one, using the constraints of the extended supersymmetry, can calculate the Kahler metric exactly. Therefore for a small enough perturbing superpotential we can hope that the previous calculation still holds, and we can find the potential on the moduli space of the ${\cal N}=2$ theory, thus checking the existence of metastable vacua. It was shown in \cite{Ooguri:2007iu, Pastras:2007qr} that actually such metastable vacua exist for an appropriate selection of the perturbing superpotential. In addition, these vacua have also found nice geometric constructions in M-theory \cite{Mazzucato:2007ah, Marsano:2008ts}.

One natural question that arises is whether the cooling universe can settle in the metastable vacuum. Studying the original construction of metastable vacua in ${\cal N}=1$ theories, it was shown that indeed it appears that at high temperatures entropy dominates energy, making the metastable vacuum globally favorable \cite{Abel:2006cr, Abel:2006my, Kaplunovsky:2007vd}. As the universe cools down, the sypersymmetric vacuum becomes globally favorable but tunneling to it is improbable since the decay rate towards it is already too small.

In this paper we examine whether a cooling universe can settle in the meta-stable vacua of ${\cal N}=2$ gauge theory softly broken to ${\cal N}=1$ by an appropriate superpotential. For simplicity we study only the $SU(2)$ ${\cal N}=2$ SYM theory without flavors. A major advantage here, in comparison to the ${\cal N}=1$ case, is that using the exact solution in the moduli space, provided by Seiberg-Witten techniques, we have more information about the area of the barrier between the metastable and supersymmetric vacua, and thus we are able to check whether there is indeed such a barrier for all temperatures. In ${\cal N}=1$ theories unfortunately the area of the barrier is the area where neither the electric nor the magnetic description of the theory are weakly coupled.

In order to study the above, for simplicity we first use a superpotential that generates a metastable vacuum at the origin of the moduli space. We study the relevant importance of the different contributions to the effective potential at high temperatures, to find that although the moduli fields are significantly lighter than the dyons, the dyons provide the major correction. Like in the ${\cal N}=1$ case we find that at high temperatures the metastable vacuum is the globally preferred vacuum. However, it turns out that if the superpotential term is much smaller than the strong coupling scale of the gauge theory, then the local minimum at the origin disappears at a range of temperatures, and the system rolls down to the SW supersymmetric vacuum as it cools down. Later, we construct the metastable vacuum at different locations in the moduli space, specify numerically the aforementioned minimum superpotential and find that this increases as the position of the metastable vacuum moves away from the origin. We also find that it is impossible for the universe to settle in metastable vacua outside an area that lies in the strongly coupled region of the moduli space.

\section{A metastable vacuum at the origin of the moduli space}

Although in \cite{Ooguri:2007iu} it was shown that we can construct a metastable vacuum at any point in the moduli space, we will start our analysis with a metastable vacuum at the origin of the moduli space. The reason for that is that we expect that at high enough temperatures the system will settle at the position where the classical symmetry of the theory is restored. This is exactly the origin of the moduli space. It is a natural guess that as the universe cools down it is easier to settle at a metastable vacuum if that resides at the same position as the minimum of the free energy at high temperatures. Moreover the theory has the discrete symmetry of reflections under the real and imaginary axis, which ensures that the position of the local minimum is not going to change at high temperatures, thus making our analysis simpler. Here we review the construction of such a metastable vacuum as presented in \cite{Ooguri:2007iu, Pastras:2007qr}.

\subsection{Review of the construction of the metastable vacuum}

\subsubsection{The ${\cal N}=2$ $SU(2)$ moduli space}

The field content of the ${\cal N}=2$ $SU(2)$ SYM without flavors consists of an $SU(2)$ adjoint gauge field and scalar, $A_\mu$ and $\phi$ respectively, and their fermionic partners. As the classical potential is given by
\begin{equation}
V\left( \phi  \right) = \frac{1}
{{g^2 }}Tr\left( {\left[ {\phi ,\phi ^\dag  } \right]} \right),
\end{equation}
there is a classical moduli space of vacua which consists of the commuting $\phi ,\phi ^\dag$ configurations. These clearly can be identified by a complex number multiplying the element of the Cartan subalgebra of $SU(2)$
\begin{equation}
\phi  = \frac{1}{2} \left( {\begin{array}{*{20}c}
   a & 0  \\
   0 & { - a}  \\

 \end{array} } \right).
\end{equation}
We will refer to this moduli space as the Coulomb branch. As $a$ is not a gauge invariant quantity we parametrize the vacua using
\begin{equation}
u = Tr\phi ^2  = \frac{1} {2}a^2.
\end{equation}

In a seminal paper \cite{Seiberg:1994aj} Seiberg and Witten managed to calculate the full quantum low energy effective theory in the Coulomb branch. It turns out that indeed the u-plane is also the quantum moduli space of the theory. Classically one would expect a singularity at $u=0$ where additional gauge fields would become massless. However there is no singularity at $u=0$ but there are singularities at $u=\pm \Lambda$, where $\Lambda$ is the strongly coupling scale of the theory. From now on we will use energy units such that $\Lambda=1$. At these singularities either a magnetic monopole or a dyon become massless. The Kahler metric on the Coulomb branch is exactly calculated in \cite{Seiberg:1994aj} and it turns out to be
\begin{equation}
  ds^2  = g(u) du d\bar u = \operatorname{Im} \left( {\tau \left( u \right)} \right)\left| {\frac{{da \left( u \right)}}
{{du}}} \right|^2 dud\bar u,
\label{eq:kahlermetric}
\end{equation}
where
\begin{equation}
\begin{gathered}
  \tau \left( u \right) = \frac{{\frac{{da_D \left( u \right)}}
{{du}}}} {{\frac{{da\left( u \right)}}
{{du}}}} \hfill \\
  a\left( u \right) = \sqrt 2 \sqrt {u + 1} {}_2F_1 \left( { - \frac{1}
{2},\frac{1}
{2};1;\frac{2}
{{u + 1}}} \right) \hfill \\
  a_D \left( u \right) = i\frac{{u - 1}}
{2}{}_2F_1 \left( {\frac{1}
{2},\frac{1}
{2};2;\frac{{1 - u}}
{2}} \right). \hfill \\
\end{gathered}
\end{equation}

\subsubsection{Softly Breaking to ${\cal N}=1$}

We now add a small superpotential term, thus breaking ${\cal N}=2$ to ${\cal N}=1$. If the superpotential is small we can assume that the Seiberg-Witten result still holds or alternatively that the Kahler metric is still given by equation \ref{eq:kahlermetric}. Then the potential in the moduli space is given by
\begin{equation}
V\left( u \right) = g^{ - 1} \left( u \right)\left|W'(u)\right|^2.
\label{scalarpotential}
\end{equation}

It was shown in \cite{Ooguri:2007iu, Pastras:2007qr} that if the superpotential has the form
\begin{equation}
W = \mu \left( u + \lambda u^3 \right)
\end{equation}
and
\begin{equation}
  \lambda _ -   < \lambda  < \lambda _ +   \hfill
\end{equation}
where
\begin{equation}
\lambda _ \pm   = \frac{1} {{24}}\left[ {1 \pm \left( {\frac{{\Gamma
\left( {\frac{3} {4}} \right)}} {{2\Gamma \left( {\frac{5} {4}}
\right)}}} \right)^4} \right],
\end{equation}
then a metastable vacuum is formed at $u=0$. In the moduli space there are also four supersymmetric vacua. Two of them lie at $u=\pm 1$, and they are there due to the singularities of the metric. In the rest on the paper we will refer to them as SW vacua. The other two lie at $u=\pm i \frac{1}{\sqrt{3\lambda}}$ and are induced by the superpotential we added. We will refer to them as W vacua. The decay rates from the metastable vacuum towards the supersymmetric vacua can become as small as desired by making $\mu$ as small as necessary.

\subsection{The Potential at High Temperatures}

The thermal contribution to the potential is given by the well known formula:
\begin{multline}
 V_{thermal}  =  - \frac{{T^4 }}{{2\pi ^2 }}\operatorname{Tr}_B \int_0^\infty  {dxx^2 \ln \left( {1 - e^{ - \sqrt {x^2  - \left( {m/T} \right)^2 } } } \right)}  \\
  + \frac{{T^4 }}{{2\pi ^2 }}\operatorname{Tr}_F \int_0^\infty  {dxx^2 \ln \left( {1 + e^{ - \sqrt {x^2  - \left( {m/T} \right)^2 } } } \right).}
\end{multline}

For $T \gg m$ the above formula can be approximated by
\begin{equation}
V_{thermal}  \simeq \frac{{T^2 }}{{24}}\operatorname{Tr}_B m^2  + \frac{{T^2 }}{{48}}\operatorname{Tr}_F m^2.
\end{equation}

For $T \ll m$ the correction to the potential is exponentially suppressed by $\frac{T}{m}$.

We would expect that at sufficiently high temperatures the thermal correction to the potential will restore the classical $SU(2)$ symmetry of the theory thus making the position that accommodates our metastable vacuum the globally prefered.

\subsubsection{The Effect of the Moduli Fields}

The first effect we need to include is the one from the moduli fields, as they are the least massive. It is interesting to check if this effect tends to restore the classical symmetry at temperatures low in comparison to the strong coupling scale of the gauge theory, as if this is the case, the effects by the other objects of the theory are going to be negligible. We thus calculate the spectrum at the metastable vacuum and at the W vacua. The gauge field and gaugino are massless in the whole moduli space. At the metastable vacuum the spectrum has already been calculated in \cite{Pastras:2007qr}. The masses of the scalars equal
\begin{equation}
\begin{gathered}
   M^2_{\phi\operatorname{Re}}  = 12\mu ^2
  {\left( \lambda -\lambda_- \right)  }   \hfill \\
   M^2_{\phi\operatorname{Im}}  = 12\mu ^2
  {\left( \lambda_+ -\lambda \right),  }   \hfill \\
\end{gathered}
\label{eq:modulimasses}
\end{equation}
while the fermion partner is massless. For simplicity we select $\lambda=\frac{1}{2}\left( \lambda_+ +\lambda_- \right)=\frac{1}{24}$. Then the masses of the two scalars become equal. We use the index $b$ for the masses of the scalars, $f$ for the fermions and $0$ to denote the position of the metastable vacuum. Then their become
\begin{equation}
\begin{array}{l}
 M_{b,0} ^2  = \frac{{\mu ^2 }}{2}\left( {\frac{{\Gamma \left( {\frac{3}{4}} \right)}}{{2\Gamma \left( {\frac{5}{4}} \right)}}} \right)^4  \\
 M_{f,0} ^2  = 0. \\
 \end{array}
\end{equation}
At the W vacuum sypersymmetry is restored, thus rendering the boson and fermion masses equal to each other and equal to
\begin{equation}
M_{b,W} ^2 = M_{f,W} ^2 = -\frac{{3\pi ^2 \mu ^2 \left| {K\left( {\frac{2}{9}\left( {1 - i2\sqrt 2 } \right)} \right)} \right|^{-2}}}{{4 {\mathop{\rm Im}\nolimits} \left( {  \frac{{\left( {\sqrt 2  - i2} \right)K\left( {\frac{1}{2} - i\sqrt 2 } \right)}}{{2K\left( {\frac{2}{9}\left( {1 - i2\sqrt 2 } \right)} \right)}}} \right)}}.
\end{equation}
Finally the energy difference between the two vacua is
\begin{equation}
\begin{array}{l}
 V_0 = \frac{4 \mu^2}{\pi }\Gamma \left( {\frac{3}{4}} \right)^4  \\
 V_W = 0.\\
 \end{array}
\end{equation}

As we assume that the superpotential is small, we can approximate the effect of the moduli by the high temperature approximation. Then the metastable vacuum becomes more favorable than the W vacuum when
\begin{equation}
V\left( 0 \right) + \frac{{M_b ^2 \left( 0 \right)}}{{12}}T_c ^2  = \frac{{M_b ^2 \left( {2\sqrt 2 i} \right)}}{8}T_c ^2.
\end{equation}
Indeed this happens for a temperature approximately equal to
\begin{equation}
T_c  \simeq 2.8\quad.
\end{equation}

The critical temperature does not depend on $\mu$ in the high temperature approximation. Unfortunately the critical temperature is larger than the strongly coupling scale of the theory. That means that the effect from other objects such as monopoles and dyons cannot be neglected. The next contributions we should account for is the one by the lightest non perturbative objects, the ones that become massless in SW vacua, namely the magnetic monopole and the $(1,1)$ dyon.

\subsubsection{The Effect of the Dyons}
\label{subsec:dyonseffect}

In order to find the effect of the dyons to the effective potential at high temperatures, we need to find the spectrum of the dyons. The spectrum of the BPS objects is calculated in \cite{Seiberg:1994rs}. The central charge equals
\begin{equation}
Z = ma + na_D,
\end{equation}
where $m$, $n$ are integers. The BPS objects saturate the BPS bound $M \ge \left| Z \right|$, thus their mass equals
\begin{equation}
M_{\left( {m,n} \right)}  = \left| {ma + na_D } \right|.
\end{equation}

\begin{figure}[h]
\begin{center}
\includegraphics[angle=0,width=0.5\textwidth]{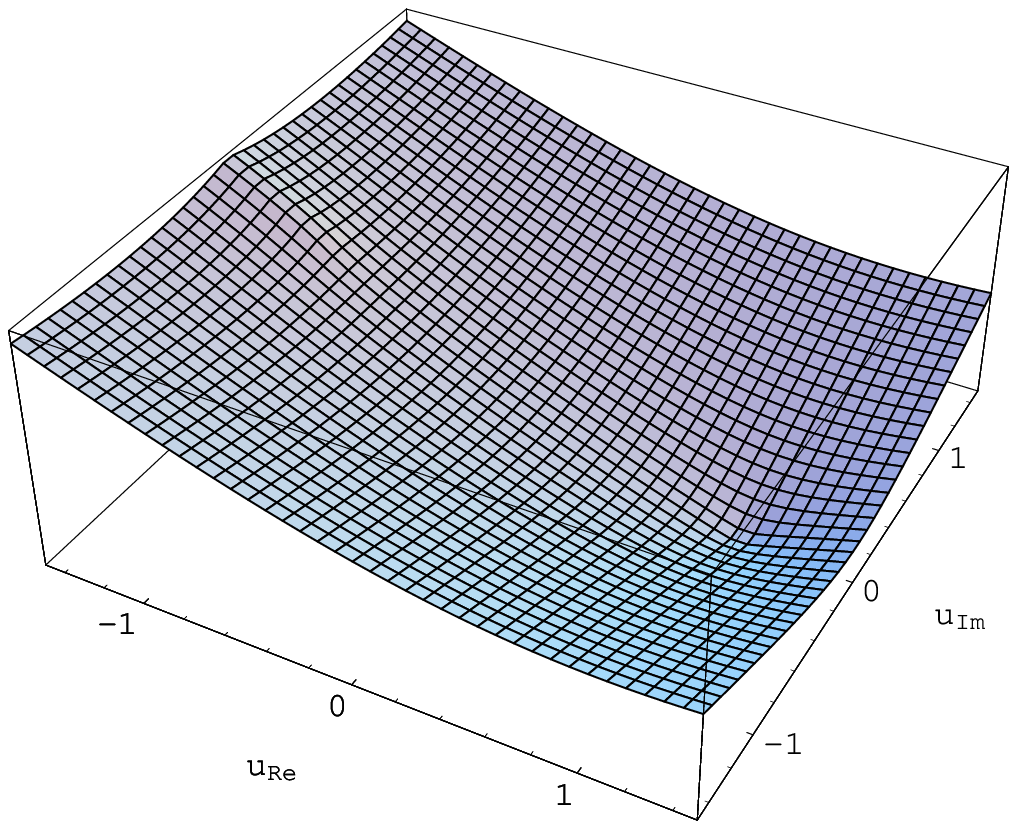}
\end{center}
\caption{The Mass of the Monopole}
\label{fig:mass01}
\begin{center}
\includegraphics[angle=0,width=0.5\textwidth]{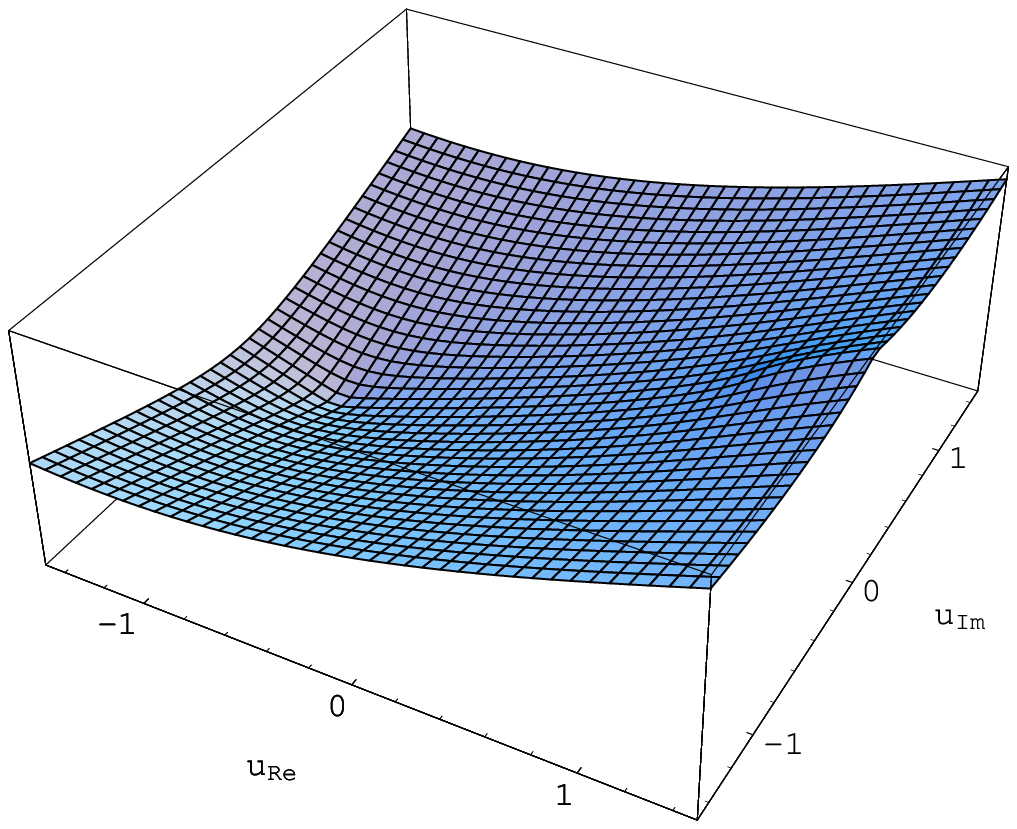}
\end{center}
\caption{The Mass of the Dyon}
\label{fig:mass11}
\end{figure}

The two lightest objects are the ones generating the singularities of the metric, the magnetic monopole and the $(1,1)$ dyon. Their masses are
\begin{equation}
\begin{array}{l}
 M_{\left( {0,1} \right)}  = \left| {a_D } \right| \\
 M_{\left( {1,1} \right)}  = \left| {a + a_D } \right|. \\
 \end{array}
\end{equation}
We plot their mass as function of the coordinate in the moduli space in figures \ref{fig:mass01} and \ref{fig:mass11}. We can notice the minima at the SW vacua, where the masses vanish.

As these non perturbative objects are actually whole ${\cal N}=2$ hypermultiplets, their contribution in the high temperature effective potential is
\begin{multline}
 V_{Dyon}  =  - \frac{{T^2 }}{{2\pi ^2 }}\left[ {4\int_0^\infty  {dxx^2 \ln \left( {1 - e^{ - \sqrt {x^2  + \left( {\left| {a_D } \right|/T} \right)^2 } } } \right)} } \right. \\
  - 4\int_0^\infty  {dxx^2 \ln \left( {1 + e^{ - \sqrt {x^2  + \left( {\left| {a_D } \right|/T} \right)^2 } } } \right)}  \\
  + 4\int_0^\infty  {dxx^2 \ln \left( {1 - e^{ - \sqrt {x^2  + \left( {\left| {a + a_D } \right|/T} \right)^2 } } } \right)}  \\
 \left. { - 4\int_0^\infty  {dxx^2 \ln \left( {1 + e^{ - \sqrt {x^2  + \left( {\left| {a + a_D } \right|/T} \right)^2 } } } \right)} } \right].
\end{multline}
The high temperature approximation of the above reads
\begin{equation}
V_{Dyon}  \simeq \frac{{T^2 }}{4}\left( {\left| {a_D } \right|^2  + \left| {a + a_D } \right|^2 } \right).
\end{equation}

\begin{figure}[h]
\begin{center}
\includegraphics[angle=0,width=0.5\textwidth]{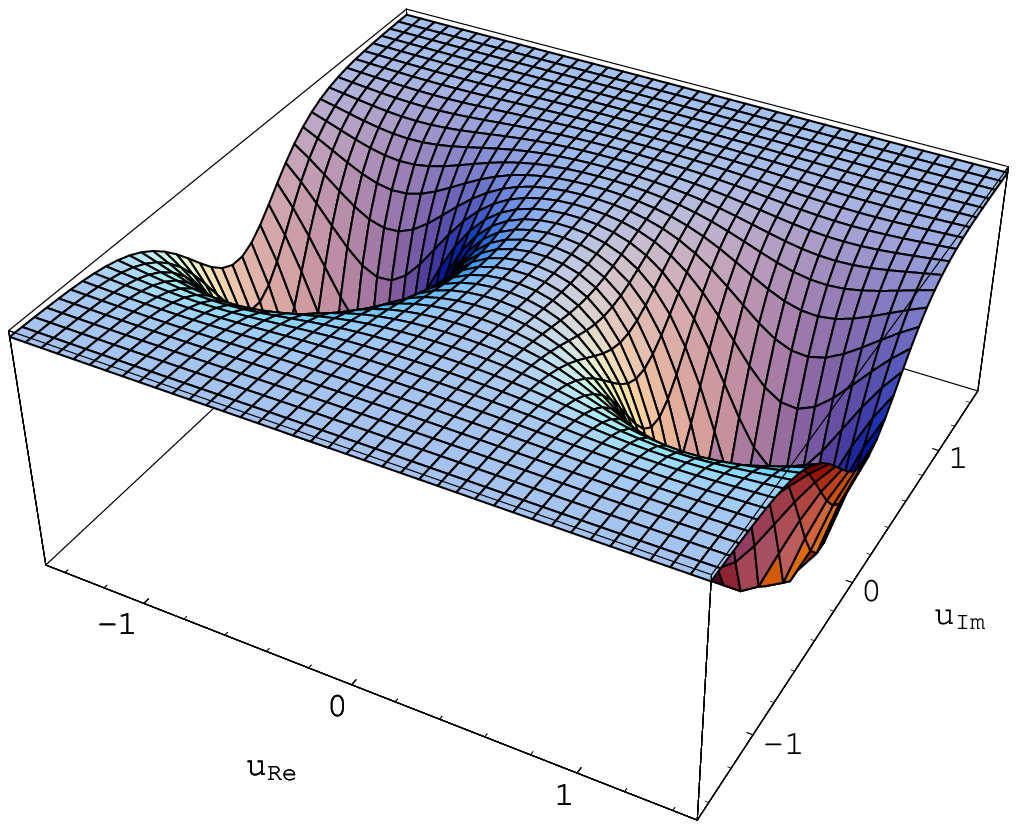}
\end{center}
\caption{The dyon contribution at low temperatures}
\label{fig:monolow}
\begin{center}
\includegraphics[angle=0,width=0.5\textwidth]{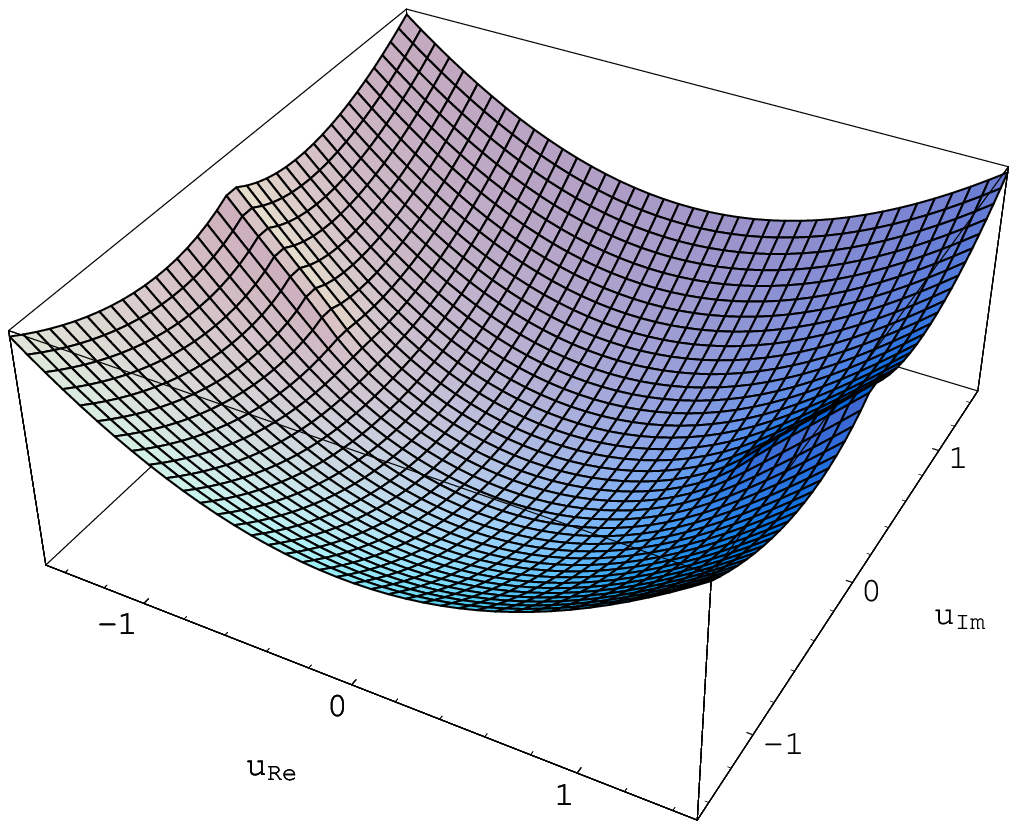}
\end{center}
\caption{The dyon contribution at high temperatures}
\label{fig:monohigh}
\end{figure}

As we can see in the graphs of the masses, between the SW vacua the dependence of the masses on $u$ is approximately linear. So we expect the sum of the squares of the masses which appears in the high temperature expansion to have a minimum at the origin. Indeed, we can see this minimum at the high temperature expansion of the dyon contribution in figure \ref{fig:monohigh}. Thus we observe that at high enough temperatures the dyon effect leads to restoration of the classical symmetry of the theory, as we expected. However the effect of the dyons at low temperatures as seen in figure \ref{fig:monolow}, leads the system to settle to one of the SW vacua.

Therefore at relatively low temperatures it is possible that the dyon effect is going to destroy the local minimum at $u=0$. In order to prevent this, the superpotential needs to be large enough so it dominates the dyon effect up to temperatures high enough so the dyon contribution looks like the one seen in figure \ref{fig:monohigh}. Therefore our demand for the universe to settle at the metastable vacuum as universe cools down constraints the hierarchy between the superpotential scale $\mu$ and the strongly coupling scale of the gauge theory $\Lambda$.

\subsubsection{Symmetry restoration}
\label{subsec:restoration}

\begin{figure}[h]
\begin{center}
\includegraphics[angle=0,width=0.5\textwidth]{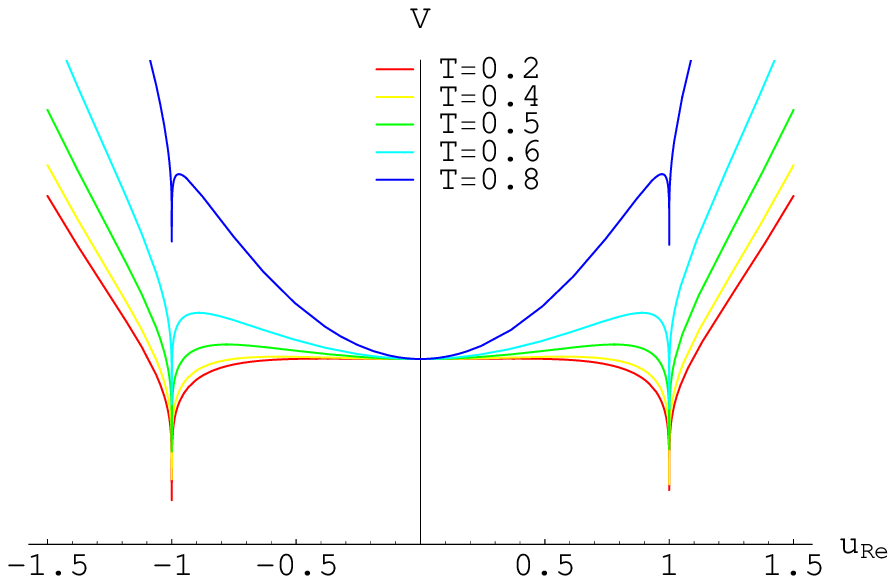}\includegraphics[angle=0,width=0.5\textwidth]{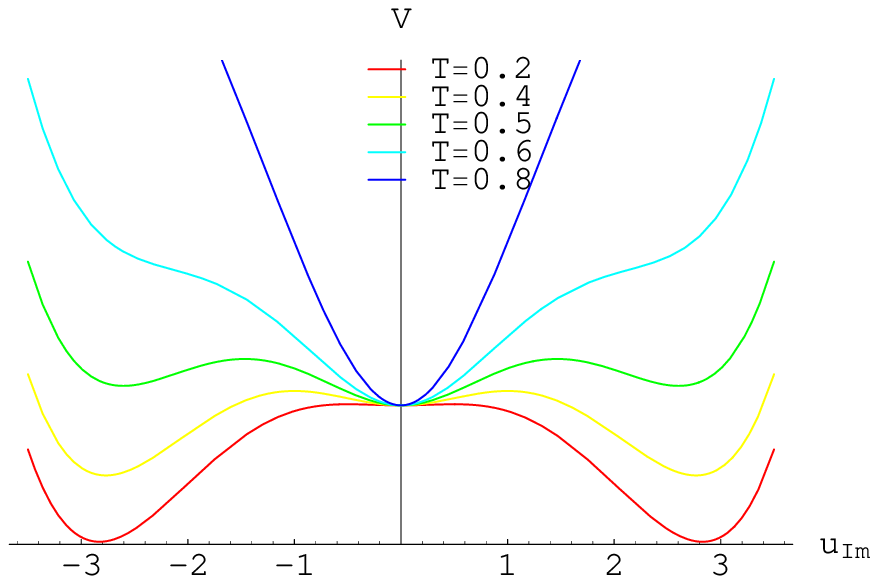}
\end{center}
\caption{The potential on the real and imaginary axes}
\label{fig:transition}
\begin{center}
\includegraphics[angle=0,width=0.5\textwidth]{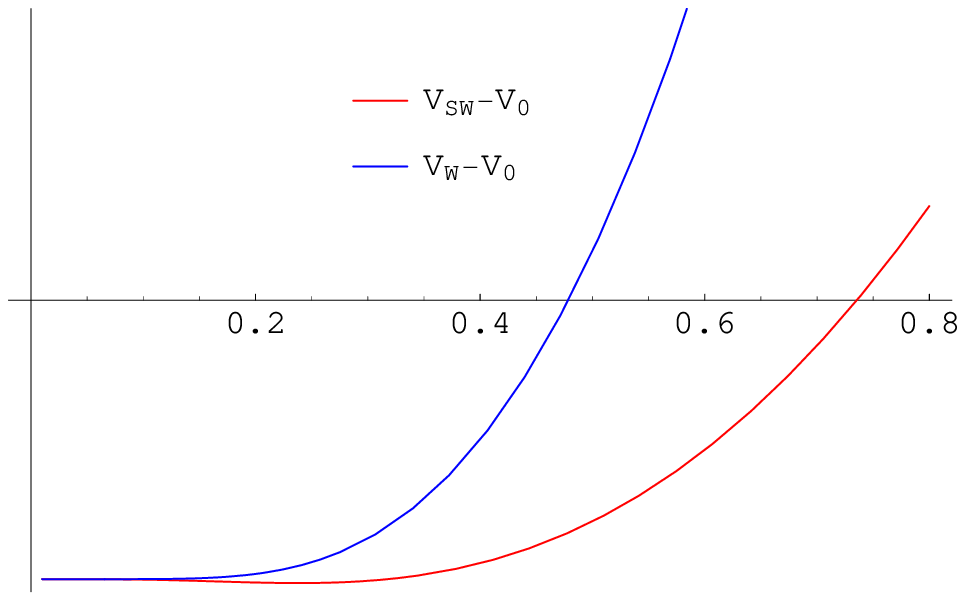}
\end{center}
\caption{The potential at the vacua}
\label{fig:potvacua}
\end{figure}

As we analyzed earlier, we expect that for high enough temperature the system will settle at the position which restores the classical symmetry of the theory, thus at $u=0$. In figures \ref{fig:transition} and \ref{fig:potvacua} we plot an example for $\lambda=\frac{1}{24}$ and $\mu=0.1$.

\begin{figure}[h]
\begin{center}
\includegraphics[angle=0,width=0.5\textwidth]{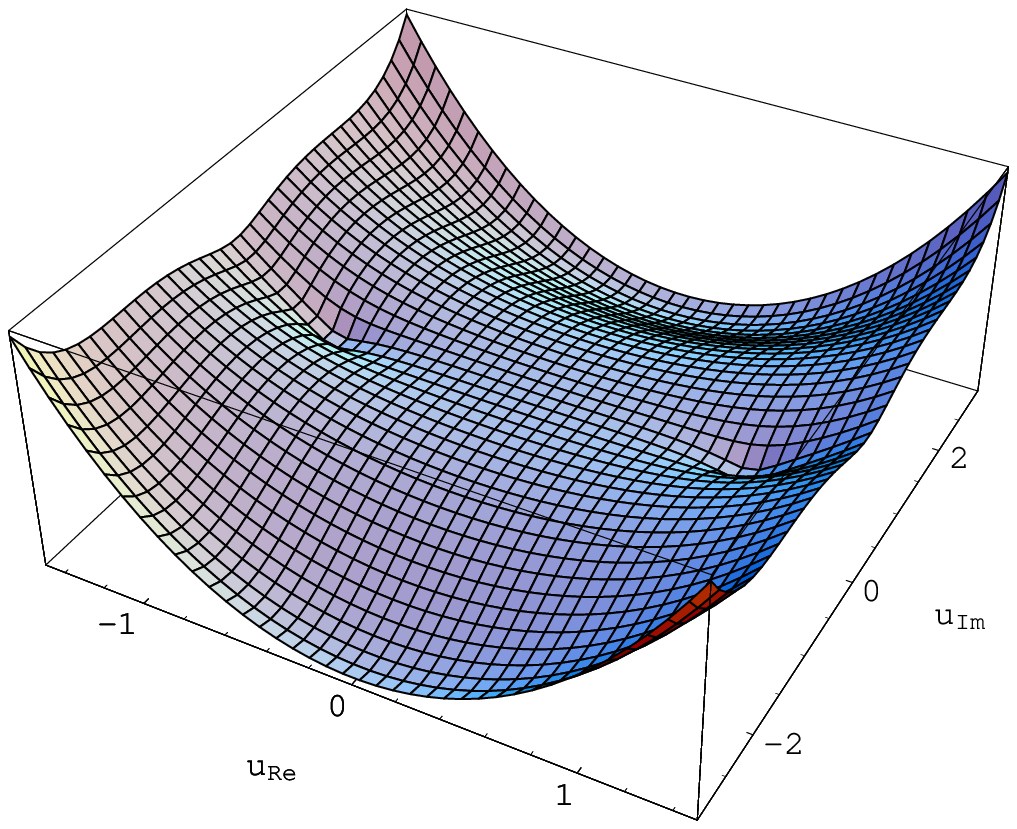}\includegraphics[angle=0,width=0.5\textwidth]{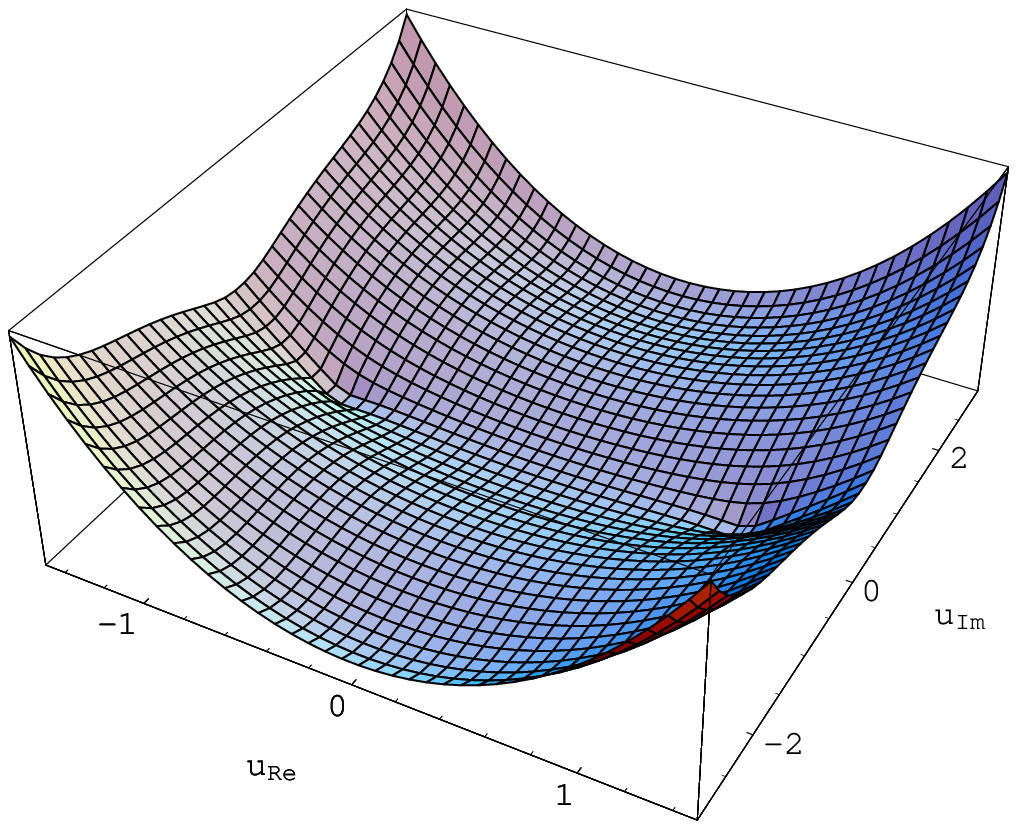}
\includegraphics[angle=0,width=0.5\textwidth]{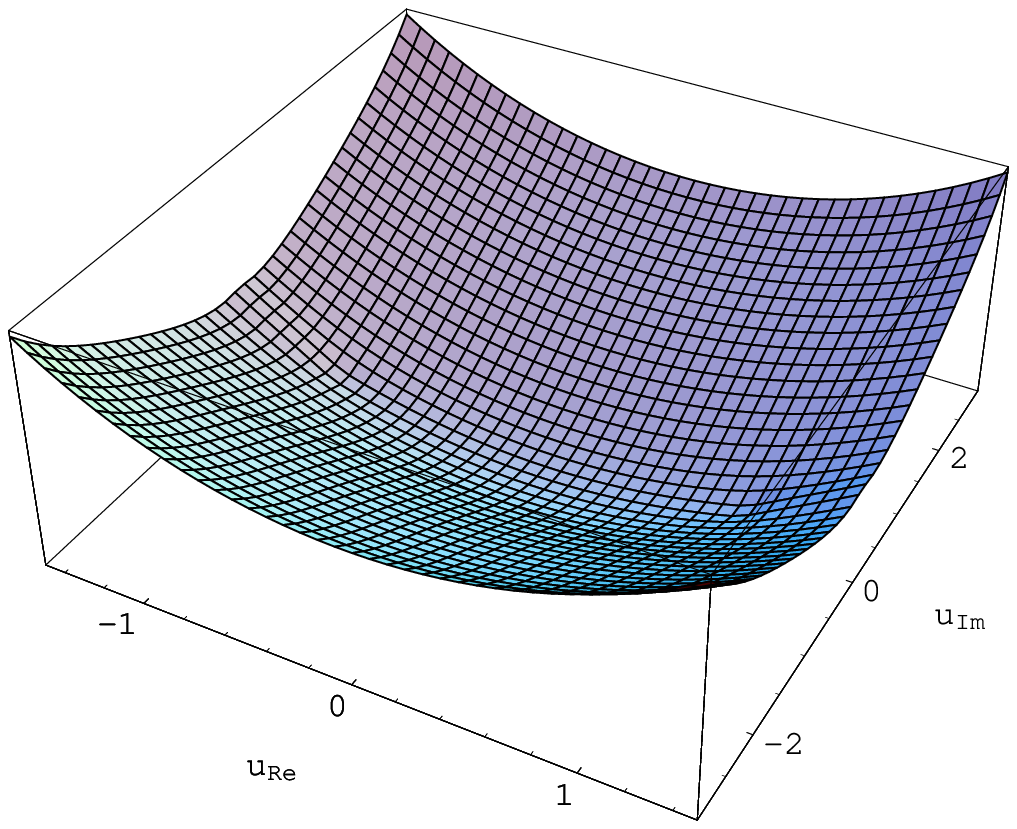}
\end{center}
\caption{The potential at $T=0.2$, $T=0.5$ and $T=0.8$}
\label{fig:potential3d}
\end{figure}

In particular in figure \ref{fig:transition} we plot the potential on the real axis for different temperatures. In figure \ref{fig:potvacua} we plot the potential difference between the metastable vacuum and the supersymmetric vacua as a function of the temperature. This plot indicates that the $u=0$ vacuum becomes the globally preferred position for temperatures above $T_c=0.65$. In figure \ref{fig:potential3d} we see the potential for several temperatures plotted in the whole moduli space.

Here we would like to make some comments. First the effect of the monopole and the dyon is much more important than the effect of the moduli fields. This can be understood by comparing the temperature where actually symmetry restoration happens with the one we calculated taking into account only the moduli fields. There are several reasons for that. The one is that as we can see in the formula for high temperature approximation, the corrections are proportional to the mass squared of the relevant field. If the phenomena we were interested in were happening at temperatures much lower than $\Lambda$, that is much lower than the typical mass of the monopole and dyon, then indeed the moduli contribution would be the most important. As this is not happening, necessarily the heavier fields contribute more. Another important factor is that the dependence of the mass of the moduli fields on the position in the moduli space is much less sharp than the dependence of the masses of the monopole and dyon. That means that even if the relevant contributions were about the same in magnitude, again the effect of the monopole and dyon would be much more important in determining which position in the moduli space is energetically favorable.

Finally we would like to point out that although in the strongly coupled region of the moduli space the monopole and the $(1,1)$ dyon are the only BPS objects present, there may be other non-BPS objects that would obviously contribute in the thermal potential as well. Since nothing is known about these objects it is not possible to compute or estimate their contributions, however it can be argued that most likely the most important contributions are the ones from the monopole and the $(1,1)$ dyon. These two objects are the only ones whose mass vanishes somewhere in the moduli space, meaning that the unknown objects have probably either much larger mass, or at least they have much less varying mass in the moduli space. In either case their effect would be less important than the effect of the monopole and $(1,1)$ dyon.

\subsection{A scan in the parameter space}

\begin{figure}[h]
\begin{center}
\includegraphics[angle=0,width=0.5\textwidth]{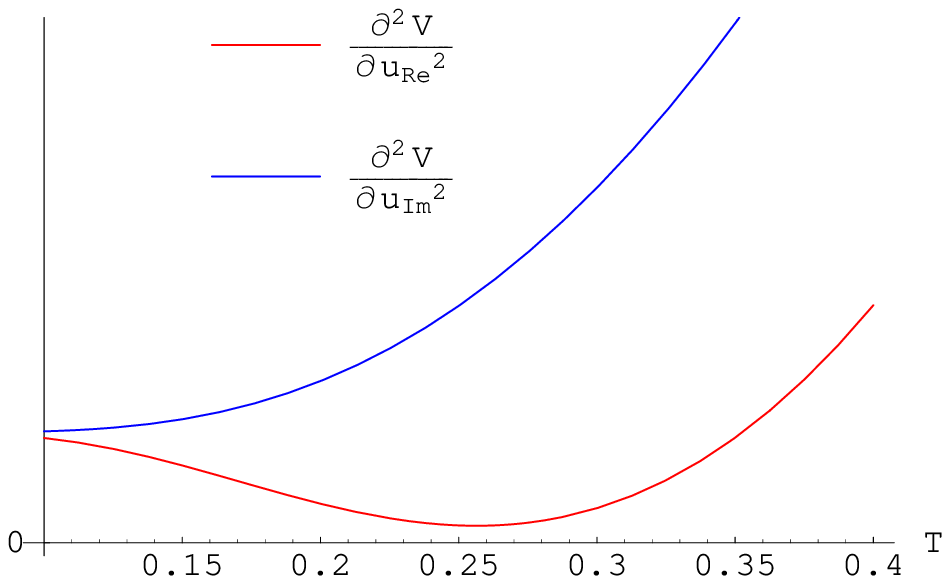}\includegraphics[angle=0,width=0.5\textwidth]{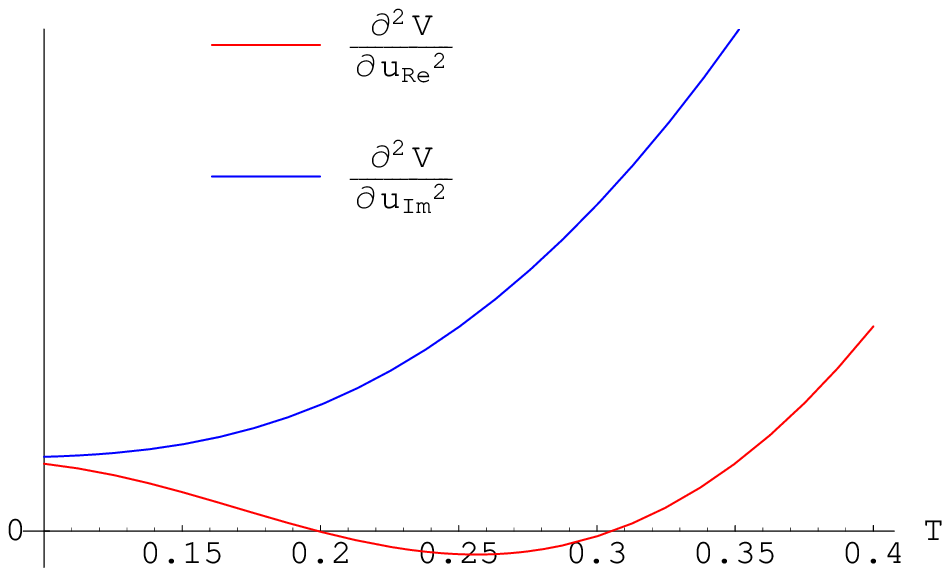}
\end{center}
\caption{The curvature of the potential for $\mu>\mu_c$ and $\mu<\mu_c$}
\label{fig:adeq}
\end{figure}

As discussed in subsection \ref{subsec:dyonseffect}, the effect of the dyons at relatively low temperatures tends to destabilize the position of the metastable vacuum. In order to have a local minimum at $u=0$ for all temperatures, it is necessary to have an adequately large superpotential. For every $\lambda$ there is a minimum $\mu_c$ that allows for the existence of the local minimum for all temperatures. In figure \ref{fig:adeq} we show the second derivatives of the potential at the metastable vacuum as functions of the temperature for adequately large $\mu$ and for insufficient $\mu$ respectively.

\begin{figure}[h]
\begin{center}
\includegraphics[angle=0,width=0.5\textwidth]{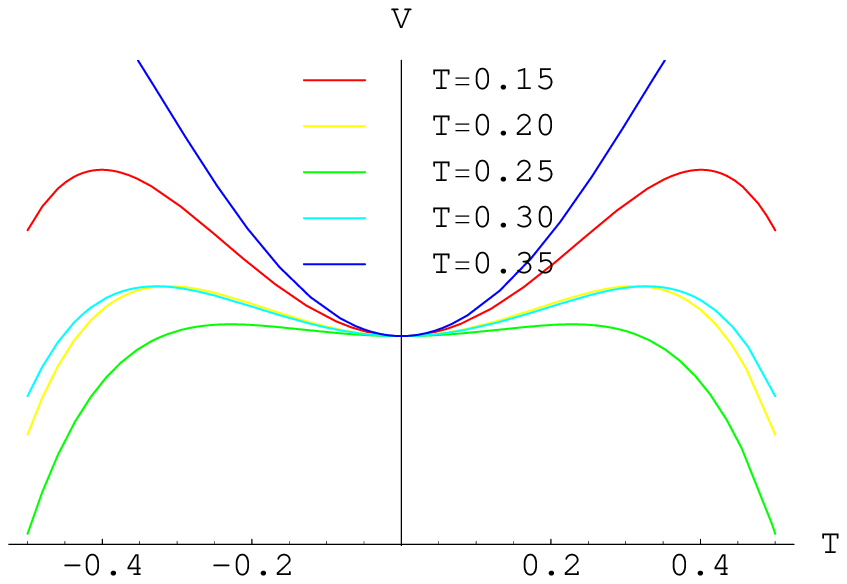}\includegraphics[angle=0,width=0.5\textwidth]{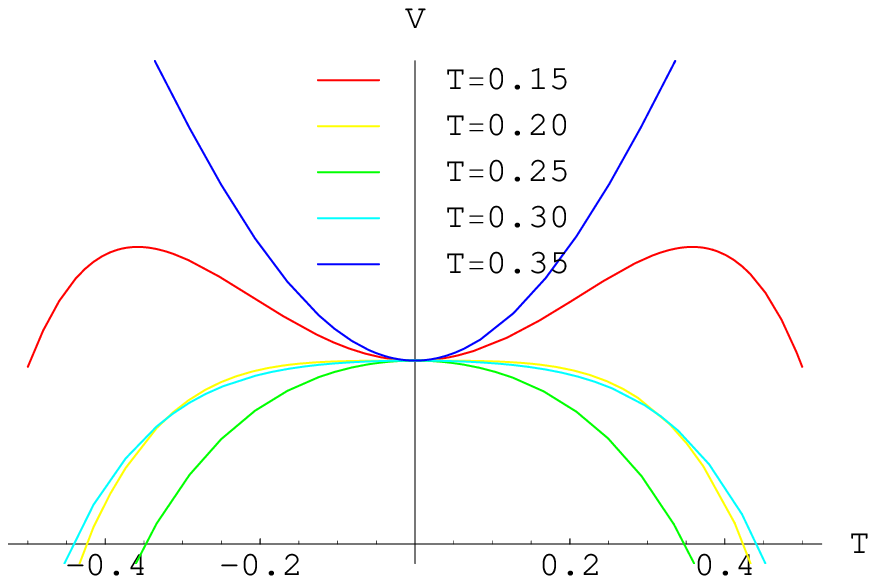}
\end{center}
\caption{The potential at the area of the metastable vacuum for $\mu>\mu_c$ and $\mu<\mu_c$}
\label{fig:vadeq}
\end{figure}

To visualize things, in figure \ref{fig:vadeq} we plot the potential on the real axis in the area of the metastable vacuum for several different temperatures, for adequately large $\mu$ and for insufficient $\mu$ respectively.

\begin{figure}[h]
\begin{center}
\includegraphics[angle=0,width=0.7\textwidth]{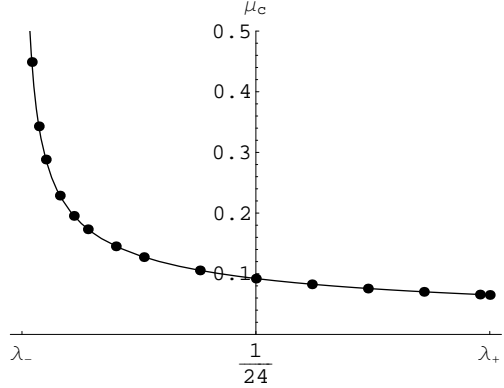}
\end{center}
\caption{$\mu_c$ as function of $\lambda$}
\label{fig:parameters}
\end{figure}

We scanned the parameter space of the superpotential, to determine the $\mu_c=\mu_c(\lambda )$ curve shown in figure \ref{fig:parameters}. The $\mu_c(\lambda )$ values are very well fitted by a curve of the form
\begin{equation}
\mu _c  = \frac{C}{{\sqrt {\lambda  - \lambda _ -  } }}.
\label{eq:paramfit}
\end{equation}

This excellent fit is not unexpected. Since as shown in subsection \ref{subsec:restoration}, the dyons and not the moduli are primarily responsible for the system behavior, the major contribution to the potential does not depend on the parameters of the superpotential. That means that we only need the second derivative of potential generated by the superpotential in the real axis to be sufficiently large to dominate the effect of the dyons. As this second derivative is given by \ref{eq:modulimasses}, it is obvious that $\mu_c$ is given by \ref{eq:paramfit}

Last, we note that for most $\lambda$, $\mu$ can be at least one order of magnitude smaller that $\Lambda$, thus making our analysis reasonably reliable.

\subsection{Decay Rates}

The supersymmetric SW vacua and the non-supersymmetric vacuum become equally favored at some temperature $T_c$ as we saw in previous sections. For temperatures larger than $T_c$ the non-supersymmetric vacuum is the preferred one, while when $T<T_c$ the supersymmetric vacuum is the preferred one. The decay rate towards the non-supersymmetric vacuum at temperatures $T>T_c$ determines whether the universe naturally results at the non-supersymmetric vacuum at high temperatures, while the decay rates towards the supersymmetric vacua at temperatures smaller than $T_c$ will determine whether the universe has a high probability to remain at the metastable non-supersymmetric vacuum as it cools down.

In order to calculate these rates, one first needs to solve the Euclidean equation of motion with the boundary condition $\phi=\phi_+$ at infinity, where $\phi_+$ is the location of the metastable vacuum. Then the decay rate equals
\begin{equation}
\Gamma = A e^{-B},
\end{equation}
where
\begin{equation}
B = S_E \left[ {\phi \left( r \right)} \right] - S_E \left[ {\phi _ +  } \right].
\label{eq:decayrate}
\end{equation}

In our case we have a thermal field theory, or else a four dimensional Euclidean theory where one dimension is compactified in a circle of circumference $\frac{1}{T}$. That means that if the bubble radius is much larger than $\frac{1}{T}$, we can approximate the four dimensional Euclidean action as
\begin{equation}
B = B_4  = S_4 \left[ {\phi \left( r \right)} \right] - S_4 \left[ {\phi _ +  } \right] \simeq \frac{{S_3 \left[ {\phi \left( r \right)} \right] - S_3 \left[ {\phi _ +  } \right]}}{T} = \frac{{B_3 }}{T}.
\end{equation}
In cases where the bubble radius is not much larger than $\frac{1}{T}$, the above approximation is not a good one, however it still serves as an upper limit for small bubble formation rates.

As the barrier between the metastable vacuum and the supersymmetric vacuum is not particularly steep, the thin wall approximation \cite{Coleman:1977py} may not be a good one. We will use the triangular approximation \cite{Duncan:1992ai}, however we need to generalize the results of the original paper in three dimensions. We do so in appendix \ref{sec:triangular}, to find that if
\begin{equation}
\frac{{\Delta \phi _ -  }}{{\Delta \phi _ +  }} \ge \frac{c}{{ - 3\left( {1 + c} \right)^{\frac{2}{3}}  + 2\left( {1 + c} \right) + 1}}
\label{eq:triangularcriterion}
\end{equation}
then
\begin{equation}
B_3 = \frac{{8\pi \left( {1 + c} \right)}}{{15}}\left( {\frac{6}{{2c + 3 - 3\left( {1 + c} \right)^{\frac{2}{3}} }}} \right)^{\frac{3}{2}} \frac{{\Delta \phi _ +  ^3 }}{{\Delta V_ +  ^{\frac{1}{2}} }},
\end{equation}
where
\begin{equation}
 \Delta \phi _ \pm =  \pm \left( {\phi _T  - \phi _ \pm  } \right) ,\quad
 \Delta V_ \pm = V_T  - V_ \pm   ,\quad
 \lambda _ \pm = \frac{{\Delta V_ \pm  }}{{\Delta \phi _ \pm }} ,\quad
 c=\frac{\lambda_-}{\lambda_+},
\end{equation}
and $\phi_+$, $\phi_-$ and $\phi_T$ are the positions of the metastable vacuum, the true vacuum and the top of the barrier between them respectively, and $V_+$, $V_-$ and $V_T$ are the relevant effective potentials. The criterion \ref{eq:triangularcriterion} is held for temperatures that are not too close to the critical temperature $T_c$. If the criterion is not true then unlike the four dimensional problem, where we can have again a formula, one has to solve numerically a set of equations in order to determine the $B$ factor, as described in the appendix.

\begin{figure}[h]
\begin{center}
\includegraphics[angle=0,width=0.5\textwidth]{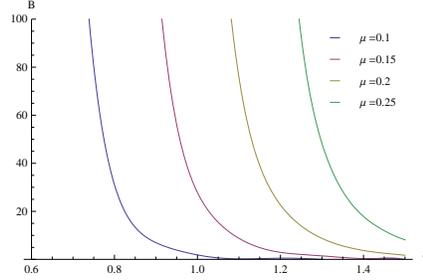}
\end{center}
\caption{$B$ for transitions from the SW vacua to the non-supersymmetric one, as function of the temperature for different choices of $\mu$}
\label{fig:bfunctionhightemp}
\end{figure}

Using the above we find the picture of figure \ref{fig:bfunctionhightemp} for the parameter $B$ describing transitions from the SW vacua towards the non-supersymmetric vacuum for temperatures larger than the critical. For simplicity we have selected $\lambda=\frac{1}{24}$.
We can see that for temperatures larger that the strongly coupled scale of the theory, the decay rates become very high. That means that if the reheating temperature is high enough then the universe is not going to remain at the SW vacua.

We will not check the decay rates for transitions from the W vacua to the non-supersymmetric one. One can see in figure \ref{fig:transition} that actually there is no barrier between these vacua at high enough temperatures. That means that provided that the reheating temperature is larger enough that the strongly coupled scale of the theory, at $T=T_c$ the universe lays at the non-supersymmetric vacuum.

\begin{figure}[h]
\begin{center}
\includegraphics[angle=0,width=0.5\textwidth]{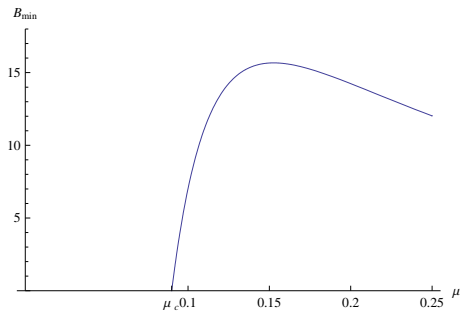}
\end{center}
\caption{The minimum value for $B$ factor as function of $\mu$}
\label{fig:Bmin}
\end{figure}

So the next step is to calculate the decay rates towards the supersymmetric vacua, for temperatures smaller than the critical. Using the triangular approximation, as we decribed above, we find the following picture.
\begin{figure}[h]
\begin{center}
\includegraphics[angle=0,width=0.5\textwidth]{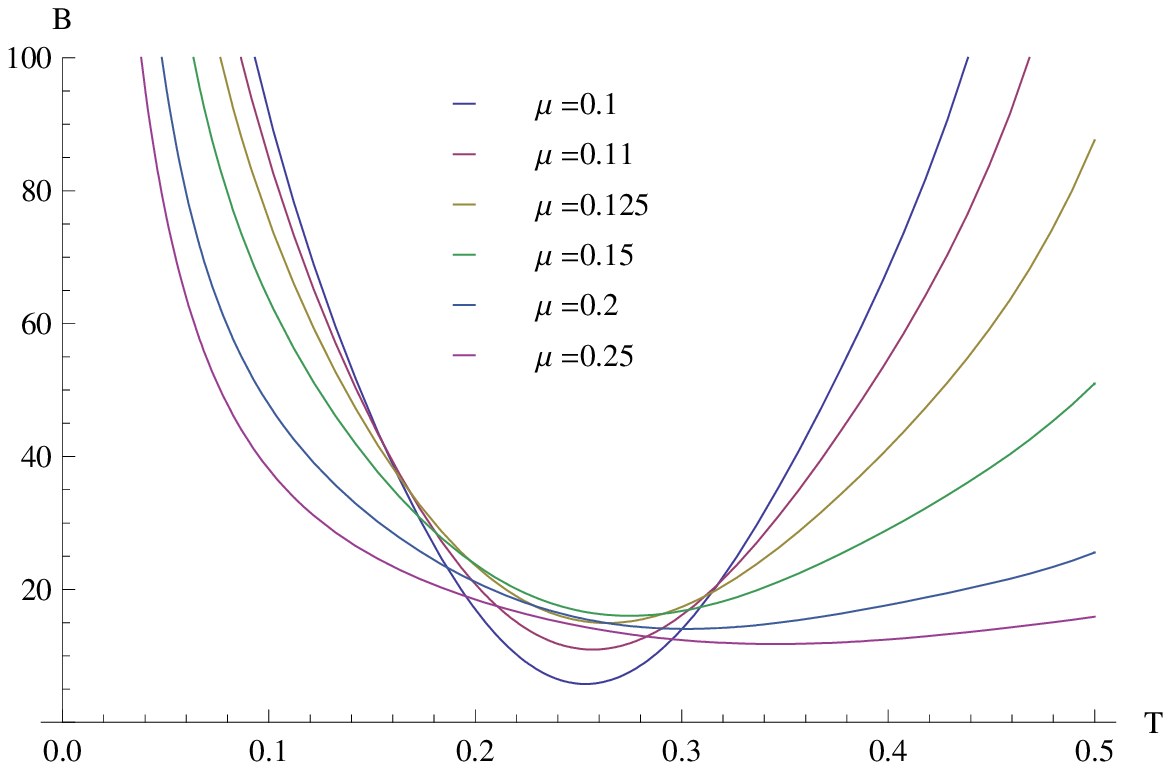}\includegraphics[angle=0,width=0.5\textwidth]{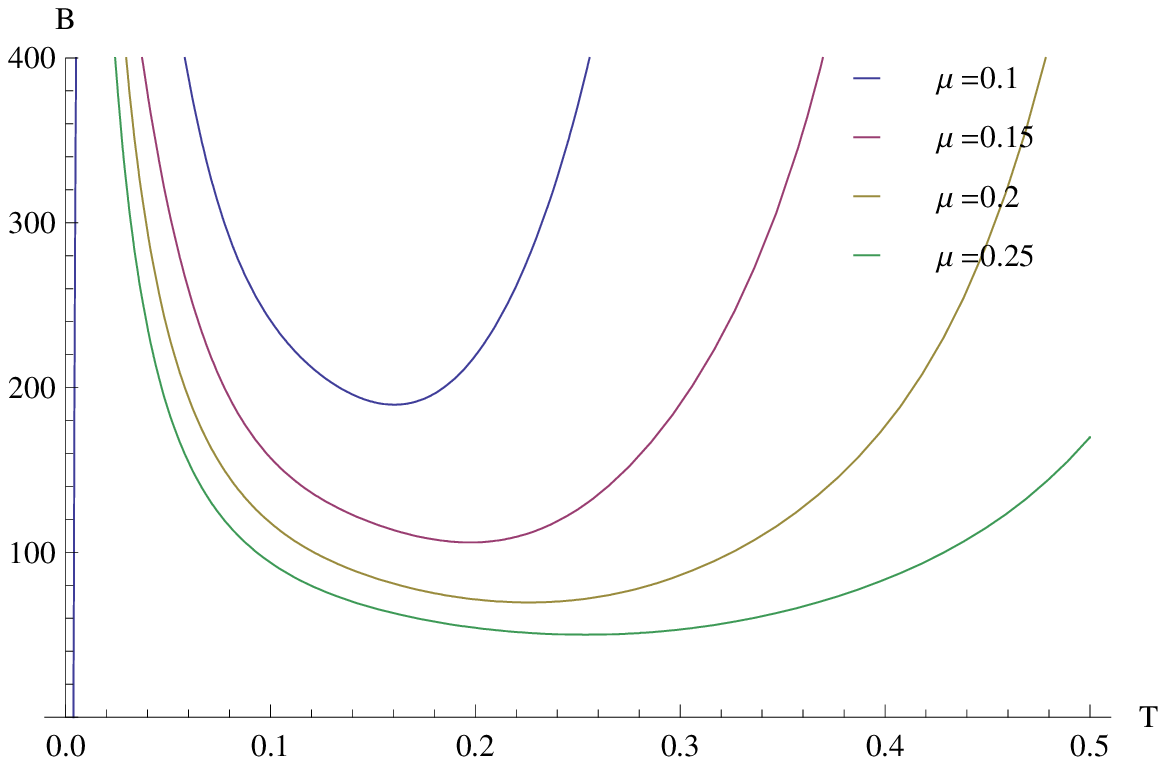}
\end{center}
\caption{$B$ for transitions from the non-supesymmetric vacuum to the SW vacua (left) and to the W vacua (right), as function of the temperature for different choices of $\mu$}
\label{fig:bfunctionlowtemp}
\end{figure}
It is clear that if the universe decays towards a supersymmetric vacuum, then most probably it decays towards a SW vacuum. We focus on this case. There are two contributions competing against each other. If $\mu$ is very large, then the barrier between the two vacua is always strong, but at the same time the potential difference between them is larger, thus increasing the probability of decay. Moreover if $\mu$ is too large the lifetime of the metastable vacuum at zero temperature may get very small. If $\mu$ is small, then the potential difference between the two vacua is smaller, but at temperatures around $T=0.25$, the barrier gets very weak (and actually if $\mu<\mu_c$ there is no barrier, as we have already seen). In figures \ref{fig:bfunctionlowtemp} and \ref{fig:Bmin} we see that highest minimum for $B$ factor is observed for $\mu$ equal to $1.5\mu_c$. So the best probability for the universe to settle down at the metastable vacuum occurs if $\mu$ is larger than $\mu_c$, but at the same order of magnitude. We would like to point that for $\mu$ about equal to $\mu_c$ according to \cite{Pastras:2007qr}, the $B$ factor for decay at $T=0$ is of the order of hundreds, thus making the metastable vacuum reasonably long living.
We would like to notice that actually in our model, even in the case of minimal decay rates that we get for $\mu$ equal to about $1.5\mu_c$, the decay rates are actually too high for the universe to remain in the supersymmetry breaking vacuum, if we assume a radiation dominated universe adiabatically cooling. However our $SU(2)$ theory is just a toy model, and it would be interesting if such calculation could be reproduced for a more realistic model based on a $\mathcal{N}=2$ softly broken to $\mathcal{N}=1$ hidden sector, and check whether this problem can be resolved.

\section{Metastable Vacua at Arbitrary Positions in the Moduli Space}

In this section we show that the results of the previous section can be generalized for metastable vacua constructed in different positions in the moduli space. It is intuitive that a metastable vacuum at the origin is more favored than the aforementioned, as the origin is the location where the universe settles at high temperatures. However we will show that there is a whole area of locations in the moduli space, in each point of which we can construct a metastable vacuum where the universe settles as it cools down, under reasonable assumptions. Thus the parameters of our theory are not so fine-tuned.

\subsection{Review of the Construction of the Metastable Vacua}

It has been shown in \cite{Ooguri:2007iu} that for every position in the moduli space there is an appropriate superpotential that generates a metastable vacuum in the specific position. If we denote this given position $u_0$, this superpotential has to be of the form
\begin{equation}
W = \mu \left[ {\left( {u - u_0 } \right) + \kappa \left( {u - u_0 } \right)^2  + \lambda \left( {u - u_0 } \right)^3 } \right].
\end{equation}

Higher order terms are irrelevant as they don't alter the curvature of the potential at $u_0$. Demanding that the superpotential has a stationary point at $u_0$ implies
\begin{equation}
\label{eq:kappa}
\kappa  = -\frac{1}{4}g\frac{{dg^{ - 1} }}{{du}},
\end{equation}
while demanding that $u_0$ is the position of a minimum gives us
\begin{equation}
\label{eq:lambdaradius}
\left| {\lambda  - \lambda _0 } \right|  < \frac{1}{{24}}g \left| {g \left| {\frac{{dg^{ - 1} }}{{du}}} \right|^2  - \frac{{d^2 g^{ - 1} }}{{dud\bar u}}} \right| \equiv r_\lambda,
\end{equation}
where
\begin{equation}
\label{eq:lambda0}
\lambda _0  = \frac{1}{{24}}g\left[ {2g\left( {\frac{{dg^{ - 1} }}{{du}}} \right)^2  - \frac{{d^2 g^{ - 1} }}{{du^2 }}} \right].
\end{equation}

So we see that we can select any $\lambda$ within a circle of given center and radius in the complex plane\footnote{In the appendix of \cite{Ooguri:2007iu} the superpotential is derived in Kahler-flat coordinates, and it is simply linear. This corresponds to the $\lambda=\lambda_0$ superpotential in our coordinates}\footnote{Note that in our analysis in previous section we considered only real $\lambda$'s. These are just a subset of the possible values for this parameter, however no new effects occur for complex values of $\lambda$}. We give more analytic formulas for the parameters using the explicit form of the $SU(2)$ metric in appendix \ref{sec:appendix}. From now on we select $\lambda=\lambda_0$ for the purposes of our analysis.

Notice that the potential generated by the above superpotential again is going to have four supersymmetric vacua, the two SW ones because of the metric, that will always lie at $u=\pm 1$ and the two because of the superpotential whose position depends on the selection of $u_0$ (and less importantly on the selection of $\lambda$) and lie at $u = u_0  + \frac{{ - \kappa  \pm \sqrt {\kappa ^2  - 3\lambda } }}{{3\lambda }}$.

\subsection{Thermal Evolution for a Metastable Vacuum Close to the Origin}

For a vacuum constructed relatively close to the origin, the physics remains approximately the same. Again the contribution to the effective potential from the monopole and the dyon is the dominant one. This effect tends to lead the system to the SW vacua at low temperatures, forcing us to demand that the superpotential is large enough for the universe to settle at the metastable vacuum as it cools down.
\begin{figure}[h]
\begin{center}
\includegraphics[angle=0,width=0.5\textwidth]{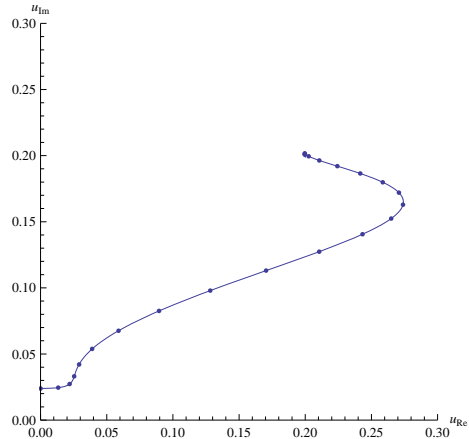}
\end{center}
\caption{The position of the minimum with temperature for $u_0=0.2+0.2i$. Temperature varies from $T=0.6$ to $T=0$ with every dot representing $\Delta T=0.025$}
\label{fig:minimumtravel}
\end{figure}
If it is not so, then the system ends up in one of the two SW vacua. As we expected, as we move away from the origin, we move away from the point where the system settles at high temperatures as well, thus making it more difficult for the universe to settle in the metastable vacuum. This forces $\mu_c$ to increase as we move away from the origin.

When the metastable vacuum is located away from the origin, another new effect happens. The discrete symmetry of the theory does not protect the position of the metastable vacuum, resulting in the local minimum moving in the moduli space as temperature changes. In figure \ref{fig:minimumtravel} the evolution of the position of the local minimum as temperature changes is displayed for $u_0=0.2+0.2i$.

\subsection{The Region of the Moduli Space that Allows Hospitable Metastable Vacua}

\begin{figure}[h]
\begin{center}
\includegraphics[angle=0,width=0.6\textwidth]{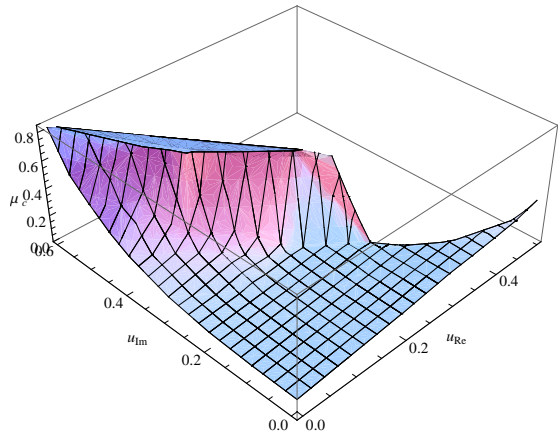}
\end{center}
\caption{The $\mu_c$ as function of $u_0$}
\label{fig:mucritical}
\end{figure}

As we already disgussed $\mu_c$ changes as we move the position of the metastable vacuum in the moduli space.
The fact that the position of the local minimum moves with temperature prevents us from making any analytical arguments for $\mu_c$. However we have performed a numerical calculation for $\mu_c$, which resulted in the plot of figure \ref{fig:mucritical}.
 We indeed see that the minimum possible $\mu_c$ occurs for $u_0=0$. This increases linearly as we move in the area close to the origin. As we move further away from the origin $\mu_c$ diverges and the larger the real part of $u_0$ the steeper the divergence. At some specific $u_{\textrm{Re}}$ the divergence becomes infinitely steep. This happens as for larger $u_{\textrm{Im}}$ than the one where the divergence occurs, the system actually rolls to the W vacua at much higher temperatures than the temperatures at which it started rolling to the SW vacua. This happens as for such large $u_0$, the position of the W vacua is closer to the origin, facilitating the aforementioned rolling.

\begin{figure}[h]
\begin{center}
\includegraphics[angle=0,width=0.6\textwidth]{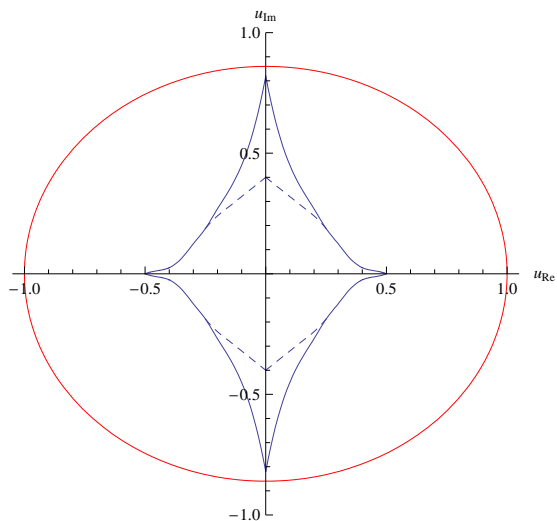}
\end{center}
\caption{The region of $u_0$ that provides hospitable vacua}
\label{fig:region}
\end{figure}

It is intuitive that if we construct the metastable vacuum too far away from the origin, the universe cannot settle there as it cools down. For example if we construct it on the real axis further away than the SW vacua, one would expect that as the universe cools down and the local minimum moves from the origin to the position of the metastable vacuum, the universe would be trapped in the SW vacuum that lies in the middle. Actually as we can already notice in figure \ref{fig:mucritical}, the constraint is much stricter than the positions of the supersymmetric vacua. In figure \ref{fig:region} the blue continuous line shows the area of the moduli space where we can construct a metastable vacuum where the universe can settle as it cools down.
The dashed line constrains us a little more demanding that the $\mu_c$ is reasonably small. The red line is the marginal stability curve, which qualitatively separates the strongly coupled region of the moduli space from the weakly coupled region. So we can see that an inhabitable metastable vacuum always lies in the strongly coupled region.

\section*{Acknowledgments}

We would like to thank D. Anninos, N. Arkani-Hamed, J. Marsano and especially K. Papadodimas for their valuable comments. E.K. would like to state that her main contribution consists of numerical calculations for sections 2.4 and 3.

\appendix

\section{Tunneling Rates in the Triangular Approximation in Arbitrary Dimensions}
\label{sec:triangular}

To calculate the decay rates from the metastable vacuum at high temperatures in the triangular approximation, we need to generalize the results of \cite{Duncan:1992ai} in three dimensions. Here we find the decay rates in arbitrary dimensions, following exactly the same procedure and notations as in the original paper.

\begin{figure}[h]
\begin{center}
\includegraphics[angle=0,width=0.7\textwidth]{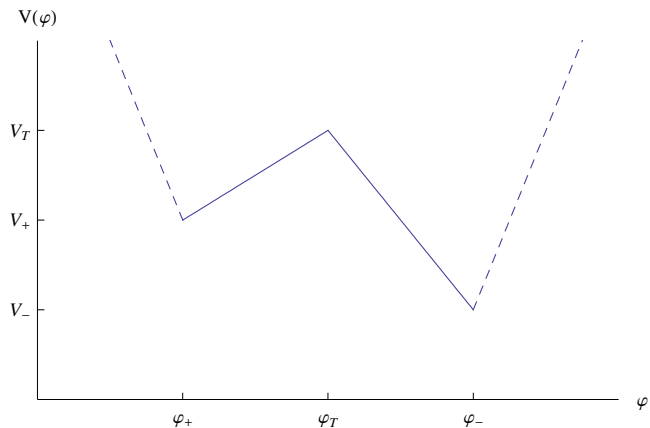}
\end{center}
\caption{The potential barrier}
\label{fig:triangular}
\end{figure}

We approximate the potential barrier by a triangular barrier like in figure \ref{fig:triangular}.
We define the position of the false vacuum as $\phi_+$, the position of the true vacuum as $\phi_-$ and the position of the maximum of the barrier as $\phi_T$. The relevant potentials are defined to be $V_+$, $V_-$ and $V_T$ respectively.

The tunneling solution is a function only of the Euclidean radius. Then the action can be reduced as
\begin{equation}
S_E \left[ \phi  \right] = V_{d-1} \int_0^\infty  {drr^{d - 1} \left[ {\frac{1}{2}\dot \phi ^2  + V\left( \phi  \right)} \right],}
\end{equation}
where $V_d R^d$ is the surface of a $d$-dimensional sphere with radius $R$. The equation of motion can be easily derived from the action:
\begin{equation}
\ddot \phi  + \frac{{d - 1}}{r}\dot \phi  = V'\left( \phi  \right),
\end{equation}
where the dot means differentiation according to $r$ and the prime means differentiation according to $\phi$.

An appropriate solution should satisfy the following boundary conditions
\begin{equation}
 \mathop {\lim }\limits_{r \to \infty } \phi \left( r \right) = \phi _ +  ,\quad
 \dot \phi \left( 0 \right) = 0.
\end{equation}
Then the tunneling rate equals
\begin{equation}
\Gamma = A e^{-B},
\end{equation}
where
\begin{equation}
B = S_E \left[ {\phi \left( r \right)} \right] - S_E \left[ {\phi _ +  } \right].
\end{equation}

We define
\begin{equation}
 \Delta \phi _ \pm =  \pm \left( {\phi _T  - \phi _ \pm  } \right) ,\quad
 \Delta V_ \pm = V_T  - V_ \pm   ,\quad
 \lambda _ \pm = \frac{{\Delta V_ \pm  }}{{\Delta \phi _ \pm }} .
\end{equation}
Then the derivative of the potential reads
\begin{equation}
V'\left( \phi  \right) =  \pm \lambda _ \pm .
\end{equation}

We can argue that the field will acquire the false vacuum value at a finite radius $R_+$. Then the first boundary condition can be written as
\begin{equation}
 \phi \left( {R_ +  } \right) = \phi _ +   ,\quad
 \dot \phi \left( {R_ +  } \right) = 0.
\end{equation}

There are two possibilities for the second boundary condition. Either the field acquires the true vacuum value at some finite radius $R_-$ or it starts rolling immediately at $r=0$. We assume that the second case is true, and then the second boundary condition can be written as
\begin{equation}
 \phi \left( 0 \right) = \phi _0  ,\quad
 \dot \phi \left( 0 \right) = 0 .
\end{equation}

We now proceed to solve the equation of motion with the given boundary conditions. The general solution of the equation of motion is
\begin{equation}
\phi \left( r \right) = c_1  + \frac{{c_2 }}{{r^{d - 2} }} \pm \frac{{\lambda _ \pm  }}{{2d}}r^2.
\end{equation}
We obviously have to demand that the field acquires the value $\phi_T$ at some distance $R_T$. Fitting the boundary conditions above we get:
\begin{equation}
 \phi \left( r \right) =
 \begin{cases} \phi _0  - \frac{{\lambda _ -  }}{{2d}}r^2 ,&r < R_T  \\
 \phi _ +   + \frac{{\lambda _ +  }}{{2d\left( {d - 2} \right)}}\left( { - dR_ +  ^2  + \frac{{2R_ +  ^d }}{{r^{d - 2} }} + \left( {d - 2} \right)r^2 } \right),&R_T < r < R_+  \\
 \phi_+,&r>R_+.\\
 \end{cases}
\end{equation}

Now we have to arrange the parameters of the two branches of the solution so it is continuous and smooth. The demand that it is smooth connects $R_T$ and $R_+$ like
\begin{equation}
R_ +  ^d  = \left( {1 + c} \right)R_T ^d,
\end{equation}
while demanding that the solution is continuous (and equal to $\phi_T$ at $R_T$) gives us $f_0$ and $R_T$:
\begin{equation}
\phi _0  = \phi _T  + \frac{{\lambda _ -  }}{{2d}}R_T ^2,
\label{eq:phi0}
\end{equation}
\begin{equation}
\frac{{\lambda _ +  }}{{2d\left( {d - 2} \right)}}\left( { - d\left( {1 + c} \right)^{\frac{2}{d}}  + 2\left( {1 + c} \right) + \left( {d - 2} \right)} \right)R_T ^2  = \Delta \phi _+.
\label{eq:rt}
\end{equation}

Now it is just a matter of some algebra to substitute the above solution to \ref{eq:decayrate} and find:
\begin{equation}
B = V_d \frac{{2\left( {1 + c} \right)}}{{d\left( {d + 2} \right)}}\left( {\frac{{2d\left( {d - 2} \right)}}{{ - d\left( {1 + c} \right)^{\frac{2}{d}}  + 2\left( {1 + c} \right) + \left( {d - 2} \right)}}} \right)^{\frac{d}{2}} \frac{{\Delta \phi _ +  ^d }}{{\Delta V_ +  ^{\frac{{d - 2}}{2}} }}.
\end{equation}
For $d=4$ we get the result of \cite{Duncan:1992ai}
\begin{equation}
B = \frac{{32\pi ^2 \left( {1 + c} \right)}}{3}\frac{1}{{\left( {\sqrt {1 + c}  - 1} \right)^4 }}\frac{{\Delta \phi _ +  ^4 }}{{\Delta V_ +  }},
\end{equation}
whereas in this paper we are interested in the $d=3$ result.
\begin{equation}
B = \frac{{8\pi \left( {1 + c} \right)}}{{15}}\left( {\frac{6}{{2c + 3 - 3\left( {1 + c} \right)^{\frac{2}{3}} }}} \right)^{\frac{3}{2}} \frac{{\Delta \phi _ +  ^3 }}{{\Delta V_ +  ^{\frac{1}{2}} }}.
\end{equation}

An important point to make is that our assumption that the field starts rolling immediately at $r=0$ is good only if $\phi _0  \le \phi _ -$. Using equations \ref{eq:phi0} and \ref{eq:rt} we see that this is equivalent to
\begin{equation}
\frac{{\Delta \phi _ -  }}{{\Delta \phi _ +  }} \ge \frac{{c\left( {d - 2} \right)}}{{ - d\left( {1 + c} \right)^{\frac{2}{d}}  + 2\left( {1 + c} \right) + \left( {d - 2} \right)}},
\end{equation}
which for $d=4$ reads
\begin{equation}
\frac{{\Delta \phi _ -  }}{{\Delta \phi _ +  }} \ge \frac{{2c}}{{ - 4\left( {1 + c} \right)^{\frac{1}{2}}  + 2\left( {1 + c} \right) + 2}}
\end{equation}
and for $d=3$ reads
\begin{equation}
\frac{{\Delta \phi _ -  }}{{\Delta \phi _ +  }} \ge \frac{c}{{ - 3\left( {1 + c} \right)^{\frac{2}{3}}  + 2\left( {1 + c} \right) + 1}}.
\end{equation}

If this assumption does not hold we need to assume that the field has the value $\phi_-$ up to a radius $R_-$ and solve again the equations with these different boundary conditions.
In this case the second boundary condition is written as
\begin{equation}
 \phi \left( {R_ -  } \right) = \phi _ -   ,\quad
 \dot \phi \left( {R_ -  } \right) = 0.
\end{equation}
Then fitting the solution to the boundary conditions gives us
\begin{equation}
 \phi \left( r \right) =
 \begin{cases}
 \phi_-,&r<R_-\\
 \phi _ -   - \frac{{\lambda _ -  }}{{2d\left( {d - 2} \right)}}\left( { - dR_ -  ^2  + \frac{{2R_ -  ^d }}{{r^{d - 2} }} + \left( {d - 2} \right)r^2 } \right),&R_- < r < R_T  \\
 \phi _ +   + \frac{{\lambda _ +  }}{{2d\left( {d - 2} \right)}}\left( { - dR_ +  ^2  + \frac{{2R_ +  ^d }}{{r^{d - 2} }} + \left( {d - 2} \right)r^2 } \right),&R_T < r < R_+  \\
 \phi_+,&r>R_+.\\
 \end{cases}
\end{equation}

Demanding that the solution is smooth at $R_T$ gives us
\begin{equation}
R_ +  ^d  - R_T ^d  = c\left( {R_T ^d  - R_ -  ^d } \right),
\end{equation}
while demanding that the solution is continuous at $R_T$ (and equal to $\phi_T$) gives us
\begin{equation}
\begin{array}{l}
 \Delta\phi _ -   = \frac{{\lambda _ -  }}{{2d\left( {d - 2} \right)}}\left( { - dR_ -  ^2  + \frac{{2R_ -  ^d }}{{R_T^{d - 2} }} + \left( {d - 2} \right)R_T^2 } \right)\\
 \Delta\phi _ +   = \frac{{\lambda _ +  }}{{2d\left( {d - 2} \right)}}\left( { - dR_ +  ^2  + \frac{{2R_ +  ^d }}{{R_T^{d - 2} }} + \left( {d - 2} \right)R_T^2 } \right).\\
 \end{array}
 \end{equation}
One needs to solve numerically the system of the three equations above, in order to specify $R_+$, $R_-$ and $R_T$. Then one can calculate the action and find $B$.

\section{The Appropriate Superpotential for the $SU(2)$ Theory}
\label{sec:appendix}

\begin{figure}[h]
\begin{center}
\includegraphics[angle=0,width=0.5\textwidth]{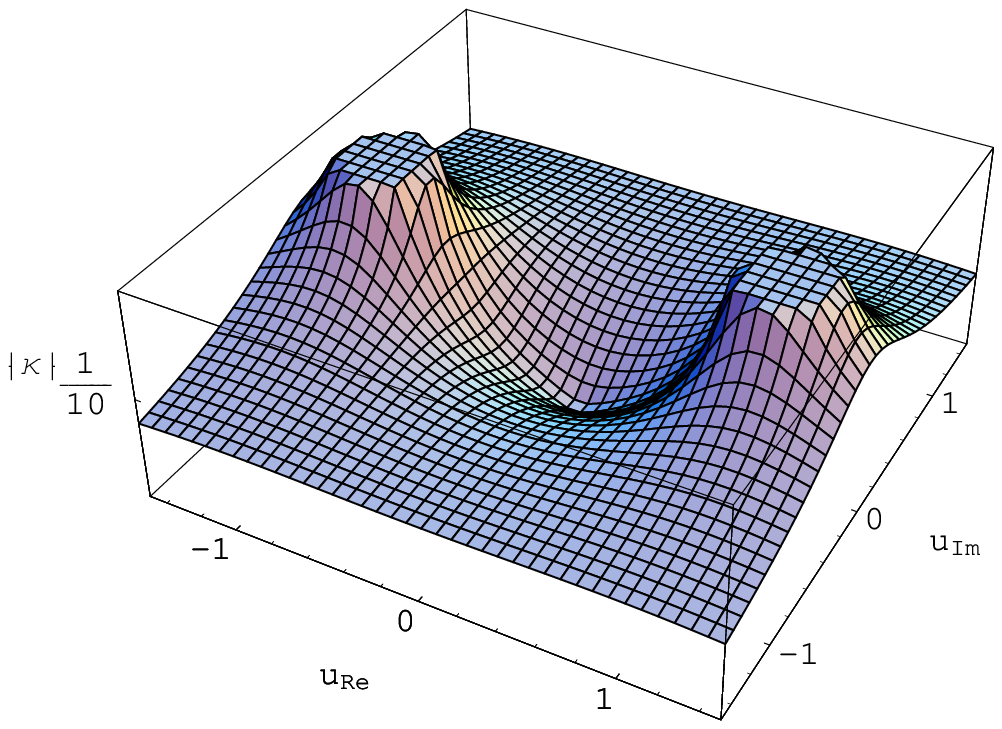}\includegraphics[angle=0,width=0.5\textwidth]{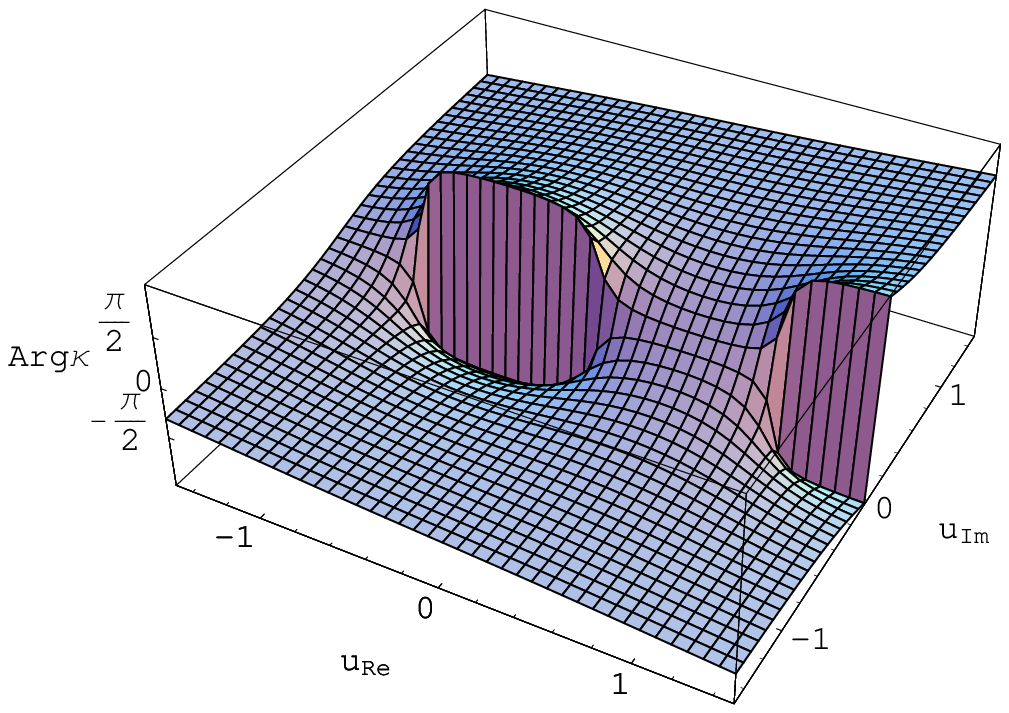}
\end{center}
\caption{The magnitude and phase of $\kappa$}
\label{fig:kappamagn}
\end{figure}

Here we present a derivation for the appropriate parameters for the superpotential, in order to construct a metastable vacuum at an arbitrary position of the moduli space of $\mathcal{N}=2$ $SU(2)$ SYM theory. First we will use the following properties of the hypergeometric functions in order to write the metric as function of elliptic integrals:
\begin{equation}
\begin{split}
 {}_2F_1 \left( { - \frac{1}{2},\frac{1}{2},1;u} \right) &= \frac{2}{\pi }E\left( u \right) \\
 {}_2F_1 \left( {\frac{1}{2},\frac{1}{2},2;u} \right) &= \frac{4}{\pi} \frac{{ {E\left( u \right) - \left( {1 - u} \right)K\left( u \right)} }}{{u}},
 \end{split}
\end{equation}
where $K\left( u \right)$ and $E\left( u \right)$ are the complete elliptic integrals of the first and second kind respectively. Then the inverse metric can be written as
\begin{equation}
g^{ - 1}  = \frac{{\pi ^2 }}{{\sqrt 2 \left| {K_2 } \right|^2 \left( {{\mathop{\rm Re}\nolimits} K_1 } \right)}},
\end{equation}
where
\begin{equation}
\begin{split}
 K_1  &\equiv \sqrt {1 + u} \frac{{K\left( {\frac{{1 - u}}{2}} \right)}}{{K\left( {\frac{2}{{1 + u}}} \right)}} \\
 K_2  &\equiv \frac{1}{{\sqrt {1 + u} }}K\left( {\frac{2}{{1 + u}}} \right).
 \end{split}
\end{equation}
In order to specify the $\kappa$ parameter we need the complex derivative of the inverse metric. Using properties of the elliptic integrals it can be written as
\begin{equation}
\frac{{dg^{ - 1} }}{{du}} = \frac{{\pi ^2 }}{{2\sqrt 2 \left| {K_2 } \right|^2 \left( {{\mathop{\rm Re}\nolimits} K_1 } \right)}}\frac{1}{{\left( {1 - u} \right)K_2 }}\left( {2E_2  + \frac{{K_1 }}{{{\mathop{\rm Re}\nolimits} K_1 }}\left( {K_2  - 2E_1  - E_2 } \right)} \right),
\end{equation}
where
\begin{equation}
\begin{split}
 E_1  &\equiv \frac{1}{{\left( {1 + u} \right)^{3/2} }}\frac{{E\left( {\frac{{1 - u}}{2}} \right)K\left( {\frac{2}{{1 + u}}} \right)}}{{K\left( {\frac{{1 - u}}{2}} \right)}} \\
 E_2  &\equiv \frac{1}{{\sqrt {1 + u} }}E\left( {\frac{2}{{1 + u}}} \right).
 \end{split}
\end{equation}
Then using equation \ref{eq:kappa}, we finally have
\begin{equation}
\kappa  = -\frac{1}{{8\left( {1 - u} \right)K_2 }}\left( {2E_2  + \frac{{K_1 }}{{{\mathop{\rm Re}\nolimits} K_1 }}\left( {K_2  - 2E_1  - E_2 } \right)} \right).
\end{equation}
In figure \ref{fig:kappamagn} we can see the magnitude and phase of the appropriate $\kappa$ as function of the position of the metastable vacuum $u_0$

\begin{figure}[h]
\begin{center}
\includegraphics[angle=0,width=0.5\textwidth]{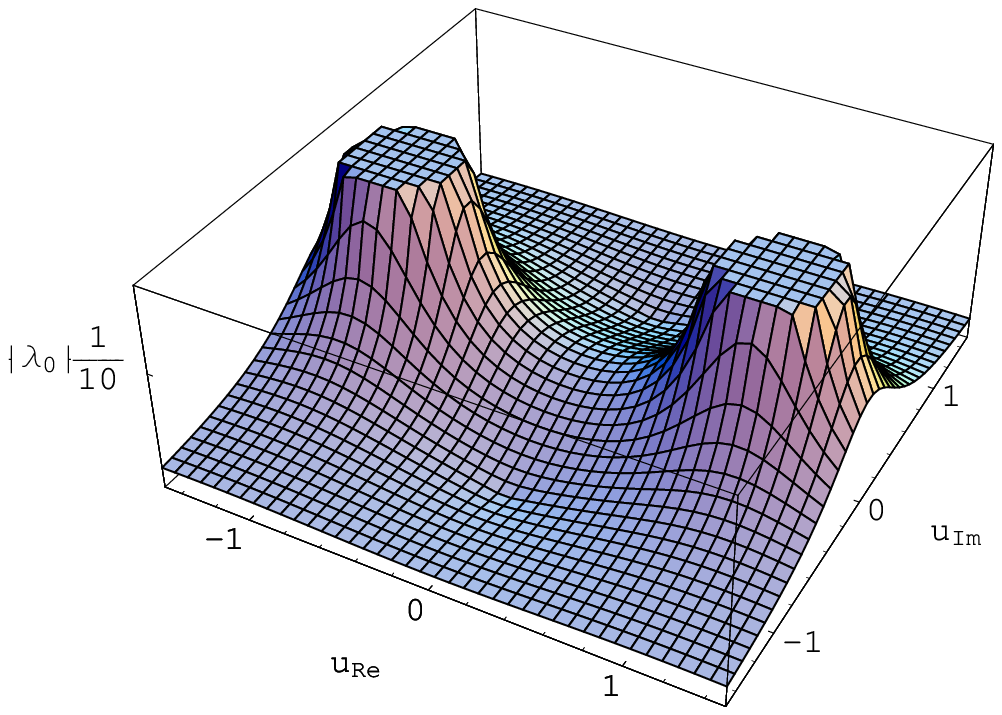}\includegraphics[angle=0,width=0.5\textwidth]{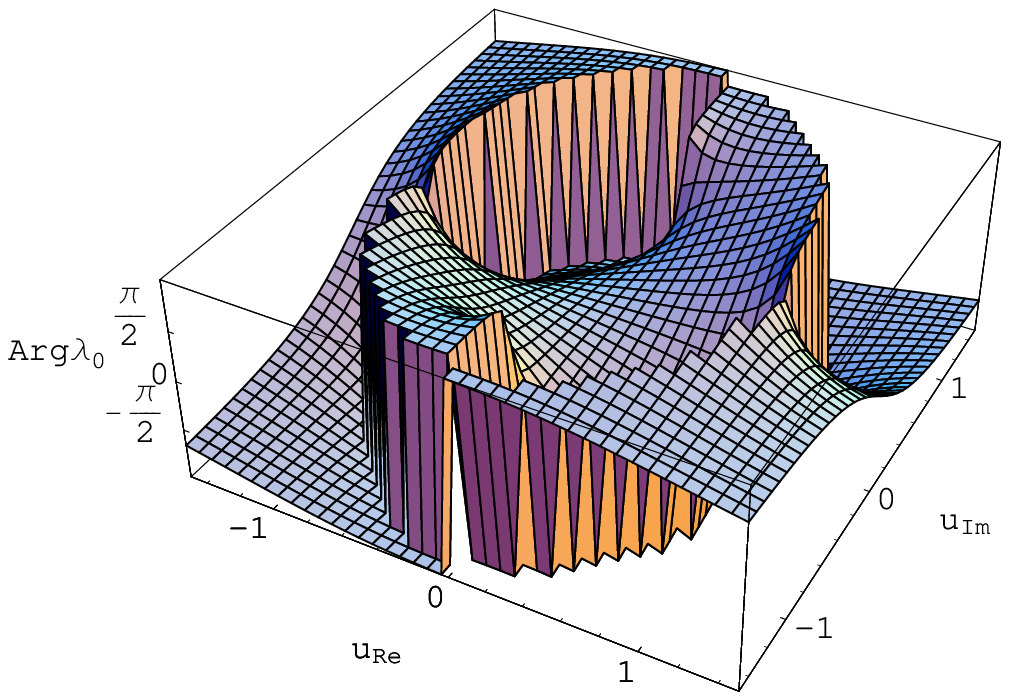}
\end{center}
\caption{The magnitude and phase of $\lambda_0$}
\label{fig:lambdamagn}
\end{figure}

In order to specify $\lambda$ we need the second derivative of the inverse metric. After some tedious algebra we get
\begin{multline}
\frac{{d^2 g^{ - 1} }}{{du^2 }} = \frac{{\pi ^2 }}{{2\sqrt 2 \left| {K_2 } \right|^2 \left( {{\mathop{\rm Re}\nolimits} K_1 } \right)}}\frac{1}{{\left( {1 - u} \right)^2 K_2 ^2 }}\left[ {\left( {2E_2  + \frac{{K_1 }}{{{\mathop{\rm Re}\nolimits} K_1 }}\left( {K_2  - 2E_1  - E_2 } \right)} \right)^2  }\right. \\ \left.{ - \frac{{4uK_2 }}{{\left( {1 + u} \right)}}\left( {2E_2  + \frac{{K_1 }}{{{\mathop{\rm Re}\nolimits} K_1 }}\left( {K_2  - 2E_1  - E_2 } \right)} \right) - \frac{{2\left( {1 - u} \right)K_2 ^2 }}{{\left( {1 + u} \right)}}} \right].
\end{multline}
Then using equation \ref{eq:lambda0} we can find the central value of the region of appropriate $\lambda$
\begin{equation}
\lambda _0  = \frac{1}{{24\left( {1 - u} \right)^2 \left( {1 + u} \right)}}\left( {\frac{{2u}}{{K_2 }}\left( {2E_2  + \frac{{K_1 }}{{{\mathop{\rm Re}\nolimits} K_1 }}\left( {K_2  - 2E_1  - E_2 } \right)} \right) + \left( {1 - u} \right)} \right),
\end{equation}
or
\begin{equation}
\lambda _0  = \frac{1}{{\left( {1 - u^2 } \right)}}\left( {\frac{{2u}}{3}\kappa  + \frac{1}{{24}}} \right).
\end{equation}
In figure \ref{fig:lambdamagn} we can see the magnitude and phase of $\lambda_0$ as function of $u_0$.

Finally in order to specify the radius of the circle of appropriate $\lambda$ we need to calculate the mixed second derivative of the inverse metric. This equals
\begin{multline}
\frac{{d^2 g^{ - 1} }}{{dud\bar u}} = \frac{{\pi ^2 }}{{2\sqrt 2 \left| {K_2 } \right|^4 \left( {{\mathop{\rm Re}\nolimits} K_1 } \right)}}\frac{1}{{\left| {1 - u } \right|^2}} \left( {2\left| {E_2 } \right|^2 }\right.\\ \left.{ + \frac{{2{\mathop{\rm Re}\nolimits} \left( {E_2 ^* K_1 \left( {K_2  - 2E_1  - E_2 } \right)} \right)}}{{\left( {{\mathop{\rm Re}\nolimits} K_1 } \right)}} + \frac{{\left| {K_1 \left( {K_2  - 2E_1  - E_2 } \right)} \right|^2 }}{{\left( {{\mathop{\rm Re}\nolimits} K_1 } \right)^2 }}} \right).
\end{multline}
We use equation \ref{eq:lambdaradius} to find
\begin{figure}[h]
\begin{center}
\includegraphics[angle=0,width=0.5\textwidth]{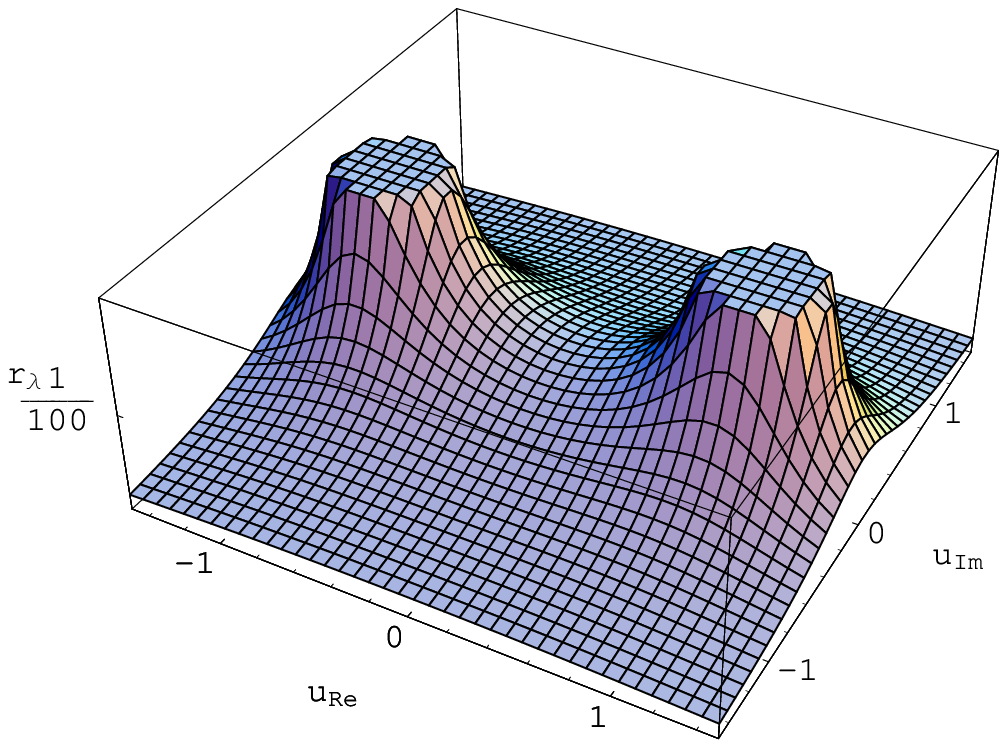}\includegraphics[angle=0,width=0.5\textwidth]{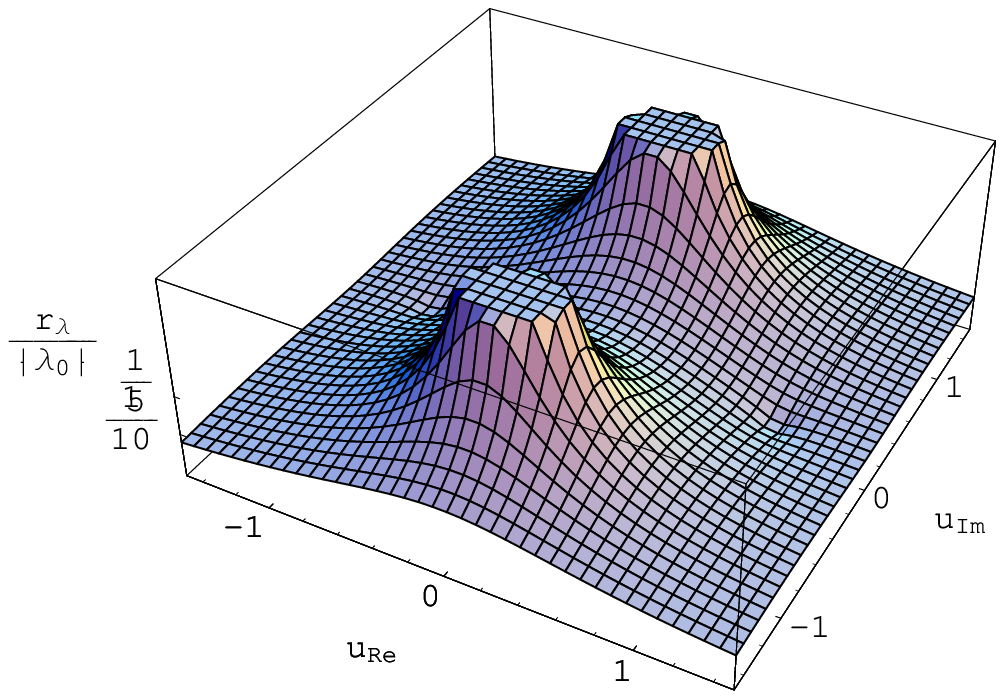}
\end{center}
\caption{The radius of the circle with appropriate $\lambda$ and the ratio of the radius to the average appropriate $\lambda$.}
\label{fig:lambdaradius}
\end{figure}
\begin{equation}
r_\lambda   = \frac{1}{{96}}\frac{1}{{\left| {1 - u} \right|^2 }}\frac{{\left| {K_1 \left( {K_2  - 2E_1  - E_2 } \right)} \right|^2 }}{{\left| {K_2 } \right|^2 \left( {{\mathop{\rm Re}\nolimits} K_1 } \right)^2 }}.
\end{equation}
In figure \ref{fig:lambdaradius} we can see $r_\lambda$ as function of $u_0$, and the ratio $\frac{r_\lambda}{\lambda_0}$, which is the most natural quantity measuring how fine tuned is the superpotential.

Finally, just for completeness, we calculate the spectrum of the moduli fields at the metastable vacuum. We can find the spectrum of the scalars diagonalizing the second derivatives of the potential. The fermion mass can be calculated directly from the superpotential. We get
\begin{equation}
\begin{array}{l}
 M_{b,u_0 } ^2  = 12\mu ^2 g^{ - 1} \left( {r_\lambda   \pm \left| {\lambda  - \lambda _0 } \right|} \right) \\
 M_{f,u_0 } ^2  = 2\mu ^2 g^{ - 1} \left| \kappa  \right|^2.  \\
 \end{array}
\end{equation}

The supertrace then equals:
\begin{equation}
\sum {\left( { - 1} \right)^F M^2 }  = 4\mu ^2 g^{ - 1} \left( {\left| \kappa  \right|^2  + 6r_\lambda  } \right),
\end{equation}
\begin{figure}[h]
\begin{center}
\includegraphics[angle=0,width=0.65\textwidth]{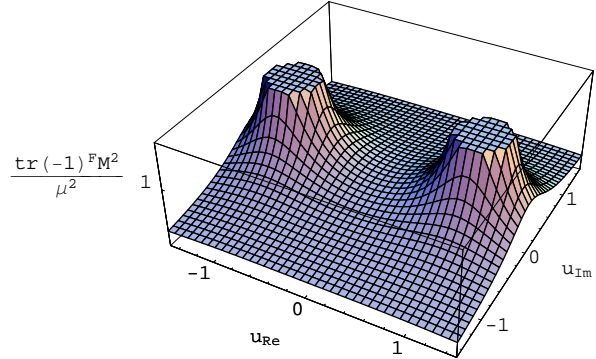}
\end{center}
\caption{The supertrace}
\label{fig:supertrace}
\end{figure}
which is independent of the selection of $\lambda$, as expected, since the supertrace depends only on the curvature of the metric and not the superpotential. In figure \ref{fig:supertrace} we see the supertrace divided with $\mu^2$.
\newline
\newline

\end{document}